\documentclass[aps,prd,groupedaddress,preprintnumbers,%
              showpacs,showkeys,floatfix]{revtex4-1}       % Documentstyle for
\usepackage[latin1]{inputenc}                    % Codepage latin1
\usepackage{graphicx}                            % Graphics package
\usepackage{epsfig}
\usepackage{epsf}
\usepackage{latexsym}                            % Load additional symbols
\usepackage{amsfonts}                            % Use AMS fonts,
\usepackage{amssymb}                             % symbols,
\usepackage{amsmath}                             % and definitions
\usepackage[mathscr]{eucal}                      % Euler font symbols
\usepackage{dcolumn}                             % Align table columns
\usepackage{multirow}
\usepackage{bm}                                  % Bold math
\usepackage{hyperref}                            % References as links
\usepackage[usenames]{color}
\usepackage{array}
\usepackage{appendix}
\usepackage{bbold}
\graphicspath{{.}{Graphics/}}                    % Path for graphics
\newcommand{\be}{\begin{equation}}
\newcommand{\ee}{\end{equation}}
\newcommand{\bea}{\begin{eqnarray}}
\newcommand{\eea}{\end{eqnarray}}
\newcommand{\tendto}{\mathop{\longrightarrow}}

\def\eq#1{Eq.~(\ref{#1})}
\def\fig#1{Fig. \ref{#1}}
\def\tbl#1{Table \ref{#1}}
\def \3{\ss }

\newcommand{\tr}{\operatorname{Tr}}

\newcommand{\beqn}{\begin{eqnarray}}
\newcommand{\eeqn}{\end{eqnarray}}

\newcommand{\idnty}{\hbox{1$\!\!$1}}

\newcommand{\reci}[1]{\frac{1}{#1}}

\newcommand{\eps}{\epsilon_{abc}}
\newcommand{\con}[3]   {\left( {#1}_a^T C\gamma_5     {#2}_b \right) {#3}_c}
\newcommand{\conm}[3]{\left( {#1}_a^T C\gamma_\mu {#2}_b \right) {#3}_c}
\newcommand{\cone}[3]{\eps \left( {#1}_a^T C\gamma_5 {#2}_b \right) {#3}_c}
\newcommand{\conme}[3]{\eps \left({#1}_a^T C\gamma_\mu {#2}_b \right) {#3}_c}

\def\cyp{a}
\def\cyi{b}

\begin{document}

\begin{titlepage}
  {\vspace{-0.5cm} \normalsize
  \hfill \parbox{60mm}{
DESY 14-096\\
                       %LPT-ORSAY 08-32 \\ 
                       %IRFU-08-29\\ 
                       % ROM2F/2008/06\\
                       %    HU-EP-08/09\\
                       %     MS-TP-08-4.\\
                        %RM3-TH/, ROM2F/2007/\\
}}\\[10mm]
  \begin{center}
    \begin{LARGE}
      \textbf{Low-lying baryon masses using  $N_f=2$ twisted mass clover-improved fermions directly at the physical point} \\
    \end{LARGE}
  \end{center}

 \vspace{.5cm}

 \vspace{-0.8cm}
  \baselineskip 20pt plus 2pt minus 2pt
  \begin{center}
    \textbf{
      C.~Alexandrou$^{(\cyp, \cyi)}$,
      C. Kallidonis$^{(\cyi)}$
}
  \end{center}

  \begin{center}
    \begin{footnotesize}
      \noindent 	
 	$^{(\cyp)}$ Department of Physics, University of Cyprus, P.O. Box 20537,
 	1678 Nicosia, Cyprus\\	
 	$^{(\cyi)}$ Computation-based Science and Technology Research Center, The Cyprus Institute, 20 Kavafi Str., Nicosia 2121, Cyprus \\
     \vspace{0.2cm}
    \end{footnotesize}
  \end{center}
  
  \begin{abstract}

The masses of the low-lying baryons are evaluated using an ensemble with two degenerate light twisted mass clover-improved quarks with mass tuned to reproduce the physical pion mass.  The Iwasaki improved gluonic action is employed. The coupling constant value corresponds to a lattice spacing of $a=0.0938(3)(2)$~fm, determined from the nucleon mass. We find that the clover term supresses isospin symmetry breaking as compared to our previous results using $N_f=2+1+1$ twisted mass fermions.
The masses of the hyperons and charmed baryons evaluated using this ensemble  are in agreement with the experimental values. We provide predictions for the mass of the doubly charmed $\Xi_{cc}^*$, as well as of the doubly and triply charmed $\Omega$s that have not yet been determined experimentally.

\begin{center}
\today
\end{center}
 \end{abstract}
\maketitle 
\end{titlepage}

%\tableofcontents
\section{Introduction}

Baryon masses of all low-lying hyperons and singly charmed baryons are well known from experiments~\cite{Agashe:2014kda}, and they therefore  serve as benchmark quantities for lattice QCD calculations. In contrast, the doubly and triply charmed sector remains mostly unexplored experimentally, though predicted by QCD and the quark model. The only available experimental evidence of doubly charmed baryons is the SELEX report of five resonances, identified as $\Xi_{ucc}^{++}(3460)$, $\Xi_{ucc}^{++}(3541)$, $\Xi_{ucc}^{++}(3780)$, $\Xi_{dcc}^{+}(3443)$ and $\Xi_{dcc}^{+}(3520)$~\cite{Mattson:2002vu,Russ:2002bw}. The $\Xi_{dcc}^+(3520)$ state was later confirmed by SELEX~\cite{Ocherashvili:2004hi,Koshkarev:2016rci}, having a mass of $3.519(2)$~GeV and an average lifetime less than 33$\cdot 10^{-15}$~s. This discovery has triggered a revival of the interest in charmed baryon spectroscopy. However, the fact that these resonances have not been confirmed by either the BABAR~\cite{Aubert:2006qw}, Belle~\cite{Chistov:2006zj,Kato:2013ynr}, LHCb at CERN~\cite{Aaij:2013voa} and FOCUS~\cite{Ratti:2003ez} experiments is somewhat puzzling. What adds to the puzzle is that theoretical studies, e.g. QCD sum rules~\cite{Wang:2010hs} as well as relativistic~\cite{Martynenko:2007je,Ebert:2002ig} and non-relativistic~\cite{Roberts:2007ni} quark models predict the $\Xi_{cc}$ mass to be 100-200~MeV higher than what SELEX has observed. This deviation is also confirmed by lattice QCD predictions, as discussed in Section~\ref{sec:comparison}. Even more interesting is the isospin splitting of about $60$~MeV between the $\Xi_{ucc}^{++}(3460)$ and the $\Xi_{dcc}^+(3520)$ states, which is one order of magnitude larger when compared to the mass differences of the other isospin partners. A possible explanation for this is that the Coulomb electro-magnetic effect dominates the strong interaction effect, hence these baryons have a very compact size~\cite{Brodsky:2011zs}. Future experimental activity on heavy baryon spectroscopy, such as the Beijing Spectrometer (BES-III)~\cite{Asner:2008nq}, the LHC~\cite{Doncheski:1995ye,Chen:2011mb,Chen:2014hqa}, the Belle-II~\cite{Aushev:2010bq} and PANDA~\cite{Lutz:2009ff} is expected to shed more light into the existence on doubly and triply charmed baryons.

Lattice QCD is in a good position to investigate the masses of doubly and triply charmed baryons using simulations with physical values of the quark masses. In view of the ongoing experimental efforts to study  charmed baryons, lattice QCD can provide valuable input. A number of lattice QCD groups have studied the ground states of spin-1/2 and spin-3/2 charmed baryons using a variety of lattice schemes, with the most recent ones using dynamical simulations~\cite{Na:2007pv,Na:2008hz,Briceno:2012wt,Liu:2009jc,Basak:2012py,Durr:2012dw,Namekawa:2013vu,Brown:2014ena,Bali:2015lka}. Many of these calculations perform chiral and continuum extrapolations. Recently, the study was extended to the higher spins of 5/2 and 7/2 and excited states using an ensemble of clover fermions on an asymmetric lattice at a pion mass of $m_\pi=391$~MeV~\cite{Padmanath:2015jea}. We make a thorough discussion of the various lattice calculations and how they compare with our results and the experimental values in Sec.~\ref{sec:comparison}.

In this work, we use an ensemble generated by the   European Twisted Mass Collaboration (ETMC) with two degenerate twisted mass clover-improved light quarks with mass tuned to reproduce the physical pion mass~\cite{Abdel-Rehim:2015pwa}.
This thus eliminates systematic uncertainties arising from chiral extrapolations. The clover-term helps in the stabilization of the simulations, while it still preserves the $\mathcal{O}(a)$ improvement of the twisted mass action~\cite{Frezzotti:2000nk,Frezzotti:2003ni} and reduces the $\mathcal{O}(a^2)$ lattice artefacts related to the breaking of the isospin symmetry. We refer to this gauge ensemble as ``the physical ensemble" from now on. This study extends our previous computations on the low-lying baryon spectrum using $N_f=2$~\cite{Alexandrou:2009qu} and $N_f=2+1+1$~\cite{Alexandrou:2014sha} twisted mass fermions using higher than physical pion masses. 

We  use Osterwalder-Seiler valence strange and charm quarks. Since our interest in this work is the baryon spectrum we
choose to use the physical mass of the
  $\Omega^-$ and $\Lambda_c^+$ baryons to tune the strange and charm quark masses, respectively. We also  opt to use the nucleon mass to fix the lattice spacing, $a$, in order to convert our lattice values to physical units. 
Comparisons of our previous results on the masses using $N_f=2$ and $N_f=2+1+1$ ensembles show no sensitivity to the dynamical  strange and charm quarks, at least within the statistical errors of the results. Therefore, as a first study using physical values of the light quark mass we will assume that strange and charm quark  unquenching effects are small.  This is also corroborated by results obtained using $N_f=2$ and $N_f=2+1+1$ twisted mass ensembles on quantities  such as the strange and charm quark masses~\cite{Blossier:2010cr,Carrasco:2014cwa} as well as the kaon and D-meson decay constants~\cite{Blossier:2009bx,Carrasco:2014poa}, which  showed no detectable unquenching effects.

Isospin breaking in the twisted mass formulation is a lattice artefact of order $a^2$. It has been shown that  adding the clover term reduces isospin splitting in the $\Delta$ multiplet~\cite{Abdel-Rehim:2015pwa} as compared to the $N_f=2+1+1$ twisted mass simulations at a similar lattice spacing. Here we study the effects of  isospin breaking effects to higher accuracy in the $\Delta$-system and in strange and charm sectors.  We compare our final results on the masses of the forty baryons studied in this work with those of  other recent lattice calculations, using a variety of discretization schemes as well as with experiment. We find remarkable agreement with  experimental results even though no continuum 
extrapolation is performed and provide predictions for the masses of doubly and triply charmed baryons.

The paper is organized as follows: The lattice action employed in the single ensemble we analyze in this work, as well as the details of the calculations, including the interpolating fields, the determination of the lattice spacing and the tuning of the strange and charm quark masses are given in Section~\ref{sec:lataction}. In Section~\ref{sec:latresults} we present our lattice results, where we study the effect of isospin symmetry breaking and discuss the various systematics. In section~\ref{sec:comparison} we compare our values with those from other lattice calculations and with experiment and in Sec.~\ref{sec:conclusions} we give our conclusions.

\section{Lattice techniques}\label{sec:lataction}

\subsection{The lattice action and simulation parameters}

In this work we analyze a  gauge ensemble produced by  ETMC at the physical pion mass~\cite{Abdel-Rehim:2015pwa}. The form of the gauge action used in the generation of this  ensemble is
\be\label{eq:gauge_action}
S_G = \beta \sum_{x;P}\left[b_0 \left(1 - \frac{1}{3}\Re\left\{{\rm Tr}\left[P^{1\times 1}(x)\right]\right\} \right) + b_1 \left(1 - \frac{1}{3}\Re\left\{{\rm Tr}\left[P^{1\times 2}(x)\right]\right\} \right) \right]\;,
\ee
where $\Re$ denotes the real part and the parameters $b_0 = 1 - 8b_1$ and $b_1 = -0.331$ are chosen such that the ``Iwasaki" improved gauge action is reproduced~\cite{Iwasaki:1983ck,Abdel-Rehim:2013yaa}. The gauge coupling parameter $\beta$ was chosen  to produce a lattice spacing of roughly $a=0.1$~fm. In the fermion sector the twisted mass fermion action for a doublet of degenerate quark flavours~\cite{Frezzotti:2000nk,Frezzotti:2003ni} is employed,  with a clover-term~\cite{Sheikholeslami:1985ij} added.
\be\label{eq:S_tml}
S_F\left[\chi,\overline{\chi},U \right]= a^4\sum_x  \overline{\chi}(x)\left(D_W[U] + m_0 + i \mu_l \gamma_5\tau^3 -\reci{4}c_{\rm SW}\sigma^{\mu\nu}\mathcal{F}^{\mu\nu}[U] \right) \chi(x)\;.
\ee
where $\tau^3$ is the third Pauli matrix acting in the flavour space, $m_0$ is the bare untwisted light quark mass, $\mu_l$ is the bare twisted light quark mass and the last term is the clover-term, with $c_{\rm SW}$ the so-called Sheikoleslami-Wohlert improvement coefficient. The field strength tensor $\mathcal{F}^{\mu\nu}[U]$ is given by~\cite{Sheikholeslami:1985ij}
\be{\label{eq:F_clover}}
\mathcal{F}^{\mu\nu}[U] = \reci{8}\left[P_{\mu,\nu}(x)+P_{\nu,-\mu}(x)+P_{-\mu,-\nu}(x)+P_{-\nu,\mu}(x) - ({\rm h.c.}) \right]
\ee
where $P_{\mu,\nu}(x)$ is a fundamental $1\times 1$ Wilson plaquette and $\sigma^{\mu\nu} = (1/2)[\gamma_\mu,\gamma_\nu]$. The value of $c_{\rm SW}$ appearing in the clover-term of \eq{eq:S_tml}  was set to $c_{\rm SW} = 1.57551$ from Pad\'{e} fits to data produced by the CP-PACS/JLQCD collaboration~\cite{Aoki:2005et}. Since the action is already ${\cal O}(a)$-improved it is not necessary to use the non-perturbative value and any value that minimizes the mass splitting between the neutral and charged pions can be used.  It was shown that using  $c_{\rm SW} = 1.57551$ reduces the isospin splitting between the neutral and charged pions to zero~\cite{Abdel-Rehim:2015pwa}. 

In \eq{eq:S_tml} $D_W[U]$ denotes the massless Wilson-Dirac operator given by 
\be \label{eq:wilson_term}
D_W[U] = \frac{1}{2} \gamma_{\mu}(\nabla_{\mu} + \nabla_{\mu}^{*})
-\frac{ar}{2} \nabla_{\mu}
\nabla^*_{\mu} 
\ee
where
\be
\nabla_\mu \psi(x)= \frac{1}{a}\biggl[U^\dagger_\mu(x)\psi(x+a\hat{\mu})-\psi(x)\biggr]
\hspace*{0.5cm} {\rm and}\hspace*{0.5cm} 
\nabla^*_{\mu}\psi(x)=-\frac{1}{a}\biggl[U_{\mu}(x-a\hat{\mu})\psi(x-a\hat{\mu})-\psi(x)\biggr]
\quad .
\ee
The quark fields denoted by $\chi$ in \eq{eq:S_tml} are in the so-called ``twisted basis". The fields in the ``physical basis", $\psi$, are obtained for maximal twist by the transformation
\be
\psi(x)=\reci{\sqrt{2}}\left(\idnty+ i \tau^3\gamma_5\right) \chi(x),\qquad
\overline{\psi}(x)=\overline{\chi}(x) \reci{\sqrt{2}}\left(\idnty + i \tau^3\gamma_5\right)
\quad.
\ee
In this paper, unless otherwise stated, the quark fields will be understood as ``physical fields", $\psi$, in particular when we define the baryonic interpolating fields.

Twisted mass fermions (TMF) provide an attractive formulation of lattice QCD allowing for automatic ${\cal O}(a)$ improvement, infrared regularization of small eigenvalues and fast dynamical simulations~\cite{Frezzotti:2003ni}. However, the $\mathcal{O}(a^2)$ lattice artefacts that the twisted mass action exhibits lead to instabilities in the numerical simulations, particularly at lower values of the quark masses and influence the phase structure of the lattice theory~\cite{Aoki:1983qi,Sharpe:1998xm,Farchioni:2004us}. The clover-term is added in the TMF action for stabilizing the simulations with quark masses low enough to reproduce the physical pion mass, while at the same time retaining automatic ${\cal O}(a)$ improvement that the TMF action features.

Maximally twisted Wilson quarks are obtained by setting the untwisted quark mass $m_0$ to its critical value $m_{\rm cr}$, while the twisted quark mass parameter $\mu$ is kept non-vanishing to give mass to the light quarks. A crucial advantage of the twisted mass formulation is the fact that, by tuning the bare untwisted quark mass $m_0$ to its critical value $m_{\rm cr}$, all physical observables are automatically 
${\cal O}(a)$ improved~\cite{Frezzotti:2003ni,Frezzotti:2003xj}. In practice, we implement maximal twist of Wilson quarks by tuning to zero the bare untwisted current quark mass, commonly called PCAC (Partially Conserved Axial Current) mass, $m_{\rm PCAC}$ \cite{Boucaud:2008xu,Frezzotti:2005gi}, which is proportional to $m_0 - m_{\rm cr}$ up to ${\cal O}(a)$ corrections. A convenient way to evaluate $m_{\rm PCAC}$ is through

\be \label{eq:m_pcac}
m_{\rm PCAC} = \lim_{t/a \gg 1} \frac{\sum_{\bf x} \langle \partial_4\tilde{A}_4^b({\bf x},t)\tilde{P}^b(0)\rangle}{\sum_{\bf x} \langle \tilde{P}^b({\bf x},t)\tilde{P}^b(0) \rangle} \qquad b=1,2 \qquad ,
\ee
where $\tilde{A}_\mu^b=\overline{\chi}\gamma_\mu\gamma_5 \frac{\tau^b}{2}\chi$ is the axial vector current and $\tilde{P}^b=\overline{\chi}\gamma_5\frac{\tau^b}{2}\chi$ is the pseudoscalar density in the twisted basis. The large $t/a$ limit is required in order to isolate the contribution of the lowest-lying charged pseudoscalar meson state in the correlators of \eq{eq:m_pcac}. This way of determining $m_{\rm PCAC}$ is equivalent to imposing on the lattice the validity of the axial Ward identity $\partial_\mu\tilde{A}_\mu^b = 2 m_{\rm PCAC} \tilde{P}^b$, $b=1,2$, between the vacuum and the charged zero three-momentum one-pion state. When $m_0$ is taken such that $m_{\rm PCAC}$ vanishes, this Ward identity expresses isospin conservation, as it becomes clear by rewriting it in the physical quark basis. 

In Table~\ref{Table:params} we list the input parameters of the calculation, namely $\beta$, $L/a$, the bare light quark mass $a\mu$ as well as the value of the pion mass. As one can see, the calculated pion mass is marginally below the physical pion mass. The value of the lattice spacing is determined from the nucleon mass, as explained in subsection~\ref{subsection:alat}. We analyze 357 gauge configurations, which provide a reasonable statistical accuracy for the observables in question.

\begin{table}[h]
\begin{center}
\renewcommand{\arraystretch}{1.2}
\renewcommand{\tabcolsep}{5.5pt}
\begin{tabular}{c|lc}
\hline\hline
\multicolumn{3}{c}{ $\beta=2.10$, $a=0.0938(3)(2)$~fm  ${r_0/a}=5.32(5)$}\\
\hline
\multirow{4}{*}{$48^3\times 96$, $L=4.5$~fm}  & $a\mu$         & 0.0009   \\
                                              & No. of Confs   & 357      \\
                              			      & $m_\pi$~(GeV)  & 0.130    \\
                        					  & $m_\pi L$      & 2.98     \\
\hline\hline
\end{tabular}
\caption{Input parameters ($\beta,L, a\mu$) of our lattice simulation with the corresponding lattice spacing ($a$), pion mass ($m_{\pi}$) as well as the number of gauge configurations analyzed.}
\label{Table:params}
\end{center}
\vspace*{-.0cm}
\end{table} 

%=============================================================
%=============================================================

\subsection{Two-point correlation functions and effective mass}

In this work we consider two-point correlation functions, defined by
\be \label{eq:two-point}
C_B^\pm (t,\vec{p}) = \sum_{\vec{x}_f}\langle\reci{4} {\rm Tr}\left(1\pm \gamma_0 \right)   \mathcal{J}_B\left(\vec{x}_f,t_f\right)\bar{\mathcal{J}}_B\left(\vec{x}_i,t_i\right)\rangle e^{-i \vec{p}\cdot\vec{x}_f}
\ee
where $\mathcal{J}_B$ is the interpolating field of the baryon state of interest  acting at the source, $\left(\vec{x}_i,t_i\right)$ and the sink, $\left(\vec{x}_f,t_f\right)$. The effective mass is obtained from the time dependence of the two-point function at $\vec{p}=0$. In this case, the symmetries of the action and the anti-periodic boundary conditions in the temporal direction for the quark fields imply that $C_B^+(t) = -C_B^-(T-t)$, where $T$ is the temporal extent of the lattice. Therefore, in order to decrease errors we average correlators in the forward and backward direction and define $C_B(t) = C_B^+(t) - C_B^-(T-t)$. In addition, the source location is chosen randomly on the whole lattice for each configuration, in order to decrease correlation among measurements. 

\begin{figure}[!ht]\vspace*{-0.2cm}
\center
\begin{minipage}{8.5cm}
{\includegraphics[width=0.9\textwidth]{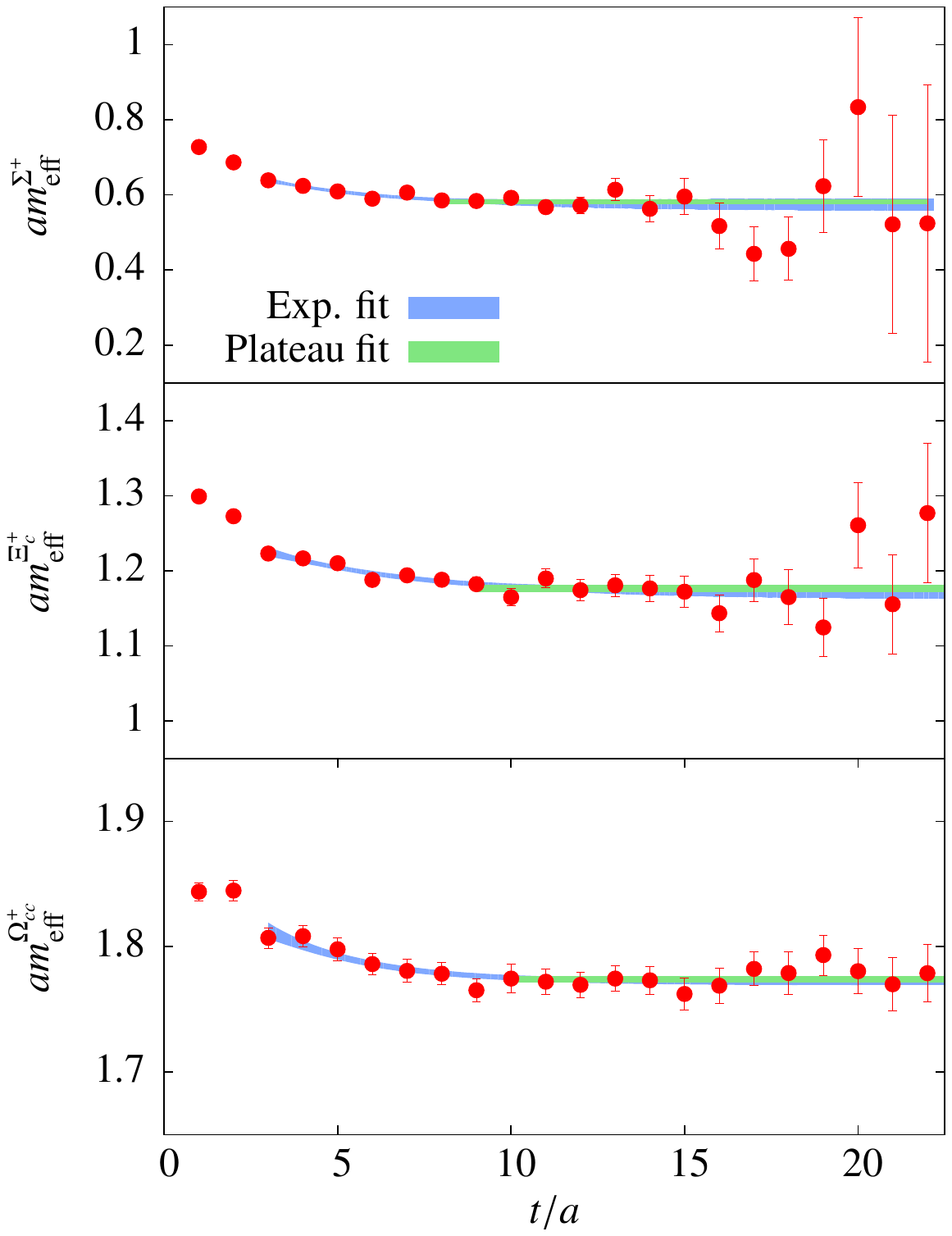}}
\end{minipage}\hfill
\begin{minipage}{8.5cm}
{\includegraphics[width=0.9\textwidth]{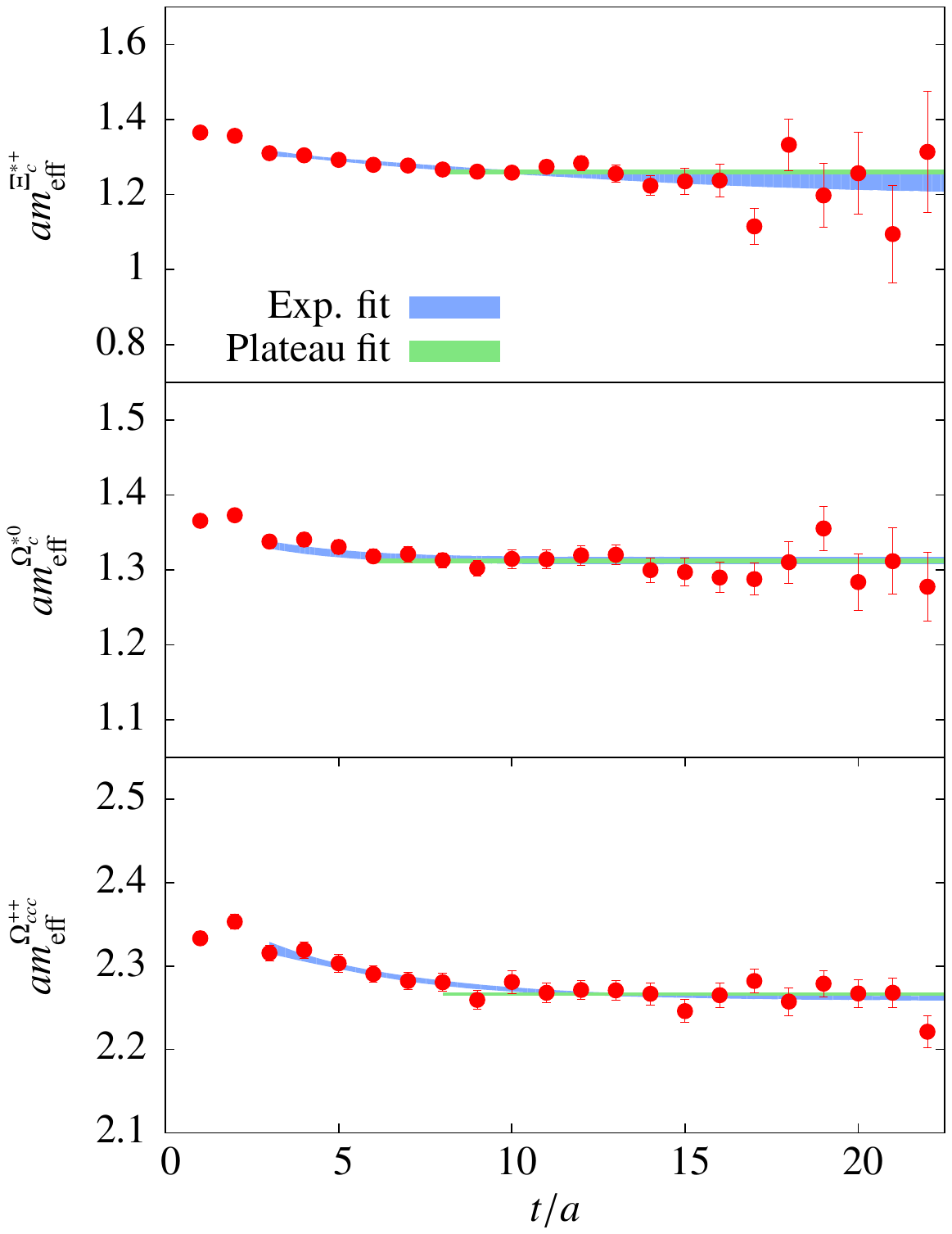}}
\end{minipage}
\caption{Representative effective mass plots for the spin-1/2 $\Sigma^+$, $\Xi_c^+$ and $\Omega_{cc}^+$ baryons (left) and for the spin-3/2 $\Xi_c^{*+}$, $\Omega_c^{*0}$ and $\Omega_{ccc}^{++}$ baryons (right). The constant fit to the plateau region is shown with the green error band and the exponential fit with the blue error band. The ground state values from the two fits are consistent.}
\label{Fig:repr_effmass}
\end{figure}

The ground state mass of a given baryon is extracted from the effective mass as
\be \label{eq:eff_mass}
am_B^{\rm eff} (t) = \log\left(\frac{C_B(t)}{C_B(t+1)} \right) = am_B + \log \left( \frac{1+\sum_{k=1}^{\infty} c_i e^{-\Delta_k t}}{1+\sum_{k=1}^{\infty} c_k e^{-\Delta_k (t+1)}} \right) \tendto_{t\rightarrow \infty} am_B
\ee
where $\Delta_k = m_k - m_B$ is the mass difference of the excited state $k$ with respect to the ground mass $m_B$. All results in this work have been extracted from correlators where Gaussian smearing is applied both at the source and sink. In general, the effective mass defined by the  correlators of a given interpolating field is  expected to have the asymptotic value. However,  applying smearing on the interpolating fields suppresses excited states, therefore yielding a plateau region at earlier source-sink time separations and thus better accuracy in the extraction of the mass. Our fitting procedure to extract $m_B$ is as follows: The sum over excited states in the effective mass given in \eq{eq:eff_mass} is truncated, keeping only the first excited state, 
\be \label{eq:eff_mass_trunc}
am_B^{\rm eff} (t) \approx am_B + \log \left( \frac{1 + c_1 e^{-\Delta_1 t}}{1+ c_1 e^{-\Delta_1 (t+1)}} \right)\;.
\ee
The upper fitting time slice boundary is kept fixed, while allowing the lower fitting time  to be  two or three time slices away from $t_i$. We then fit the effective mass to the form given in \eq{eq:eff_mass_trunc}. This exponential fit yields an estimate for $c_1$ and $\Delta_1$ as well as for the ground state mass, which we denote by $m_B^{(E)}$. Then, we perform a constant fit to the plateau region of the effective mass increasing the initial fitting time $t_1$. We denote the extracted value from the constant fit by $m_B^{(C)}(t_1)$. The final value of the mass is picked as the constant fit at the lowest $t_1$ for which the criterion
\be \label{eq:crit_1}
\Delta m_B (t_1) = | am_B^{(C)}(t_1) -  am_B^{(E)} | < \delta m_B^{(C)}(t_1)
\ee
is satisfied, where $\delta m_B^{(C)}(t_1)$ is the statistical error on $m_B^{(C)}(t_1)$. This criterion is, in most cases, in agreement with $\chi^2/{\rm d.o.f.}$ becoming less than unity.
We show representative results of the effective masses of a number of spin-1/2 and spin-3/2 baryons in~\fig{Fig:repr_effmass}, including the exponential and plateau fits. The error bands are obtained using jackknife analysis. As can be seen the exponential and plateau fits yield consistent results in the large time limit. We note that fitting directly the correlators instead of the effective masses yields compatible results.
We choose the values extracted from the constant fits on the effective masses as our final baryon masses, quoted in~\tbl{Table:phys_masses}.

%=============================================================
%=============================================================

\subsection{Interpolating fields}

In order to create baryon states on the lattice, we act on the vacuum with appropriate interpolating field operators, constructed such that they have the quantum number of the baryon of interest and reduce to the quark model wave functions in the non-relativistic limit. The forty baryons we analyze in this work consist of combinations of three out of the four quark flavors, $u$, $d$, $s$ and $c$, therefore we use SU(3) subgroups of the SU(4) symmetry to construct their interpolating fields. We use the same interpolating fields employed in our previous $N_f=2+1+1$ studies~\cite{Alexandrou:2016xok,Alexandrou:2014sha,Alexandrou:2014hsa}. For completeness we summarize below the constructions of the interpolating fields. 

In general, the interpolating fields of baryons can be written as a sum of terms of the form $\epsilon_{abc}\left[(q_1)_a^T\Gamma^A (q_2)_b\right]\Gamma^B (q_3)_c$, apart from overall  constants. The structures $\Gamma^A$ and $\Gamma^B$ are such that they give rise to the quantum numbers of the baryon state of interest. For spin-1/2 baryons, we will use the combination $(\Gamma^A,\Gamma^B)=(C\gamma_5,\mathbb{1})$ and for spin-3/2 baryons we will use $(\Gamma^A,\Gamma^B)=(C\gamma_j,\mathbb{1})$, with  $j=1,\ldots,3$. $C$ is the charge conjugation matrix.

\begin{figure}[!ht]
\begin{minipage}[t]{0.35\linewidth}
\includegraphics[width=\linewidth]{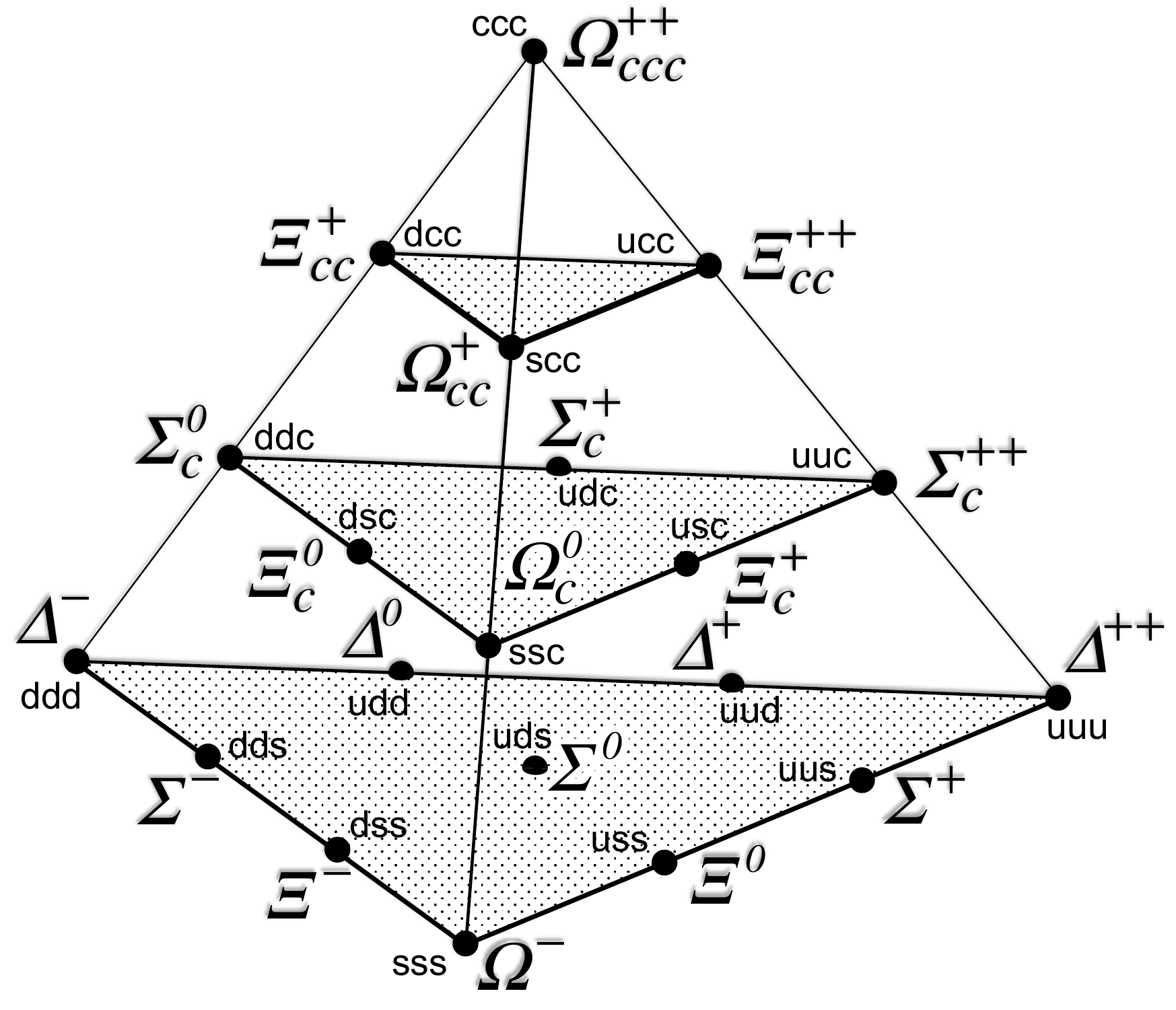}
\end{minipage} \hfill
\begin{minipage}[t]{0.35\linewidth}
\includegraphics[width=\linewidth]{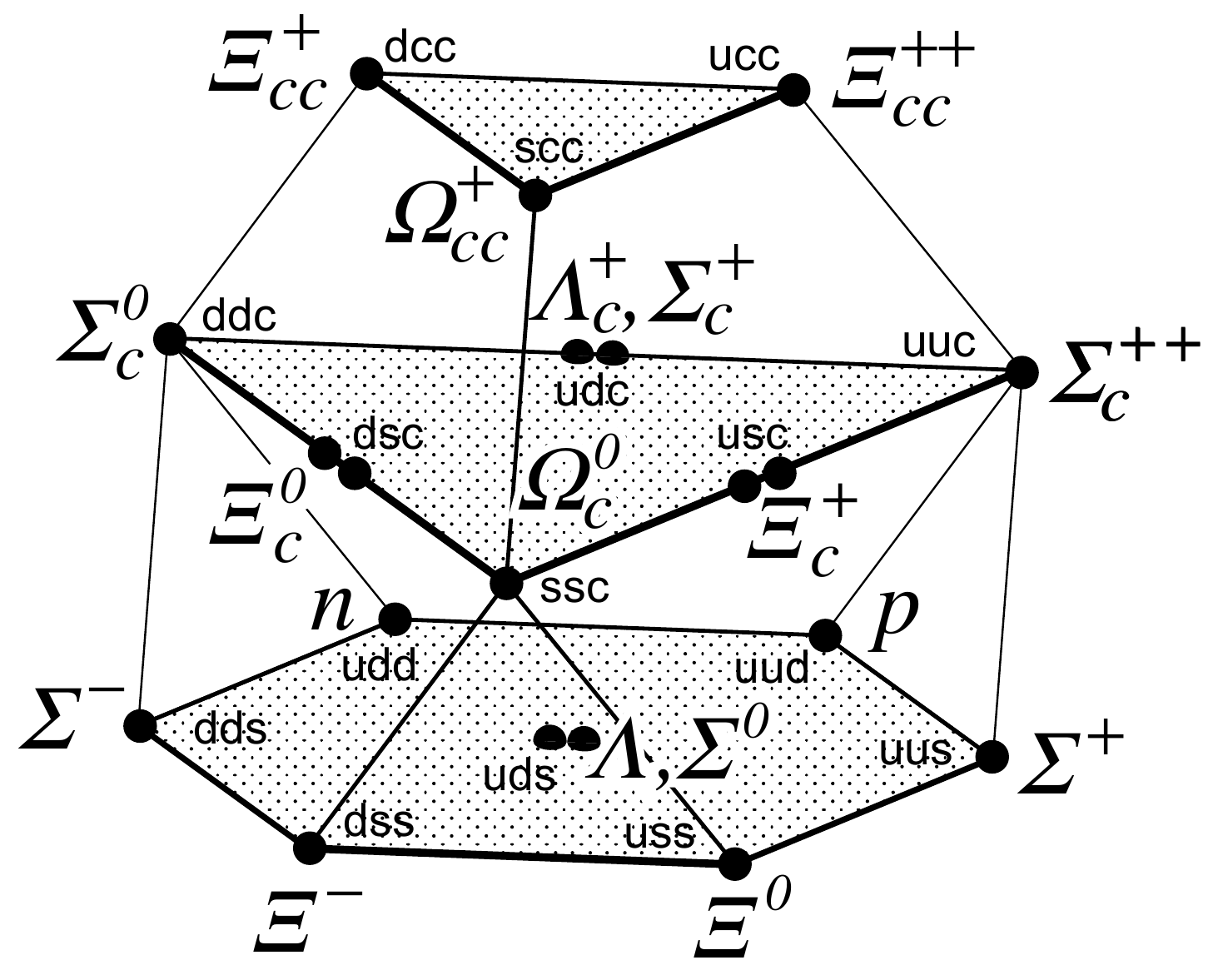}
\end{minipage} \hfill
\begin{minipage}[t]{0.28\linewidth}
\includegraphics[width=\linewidth]{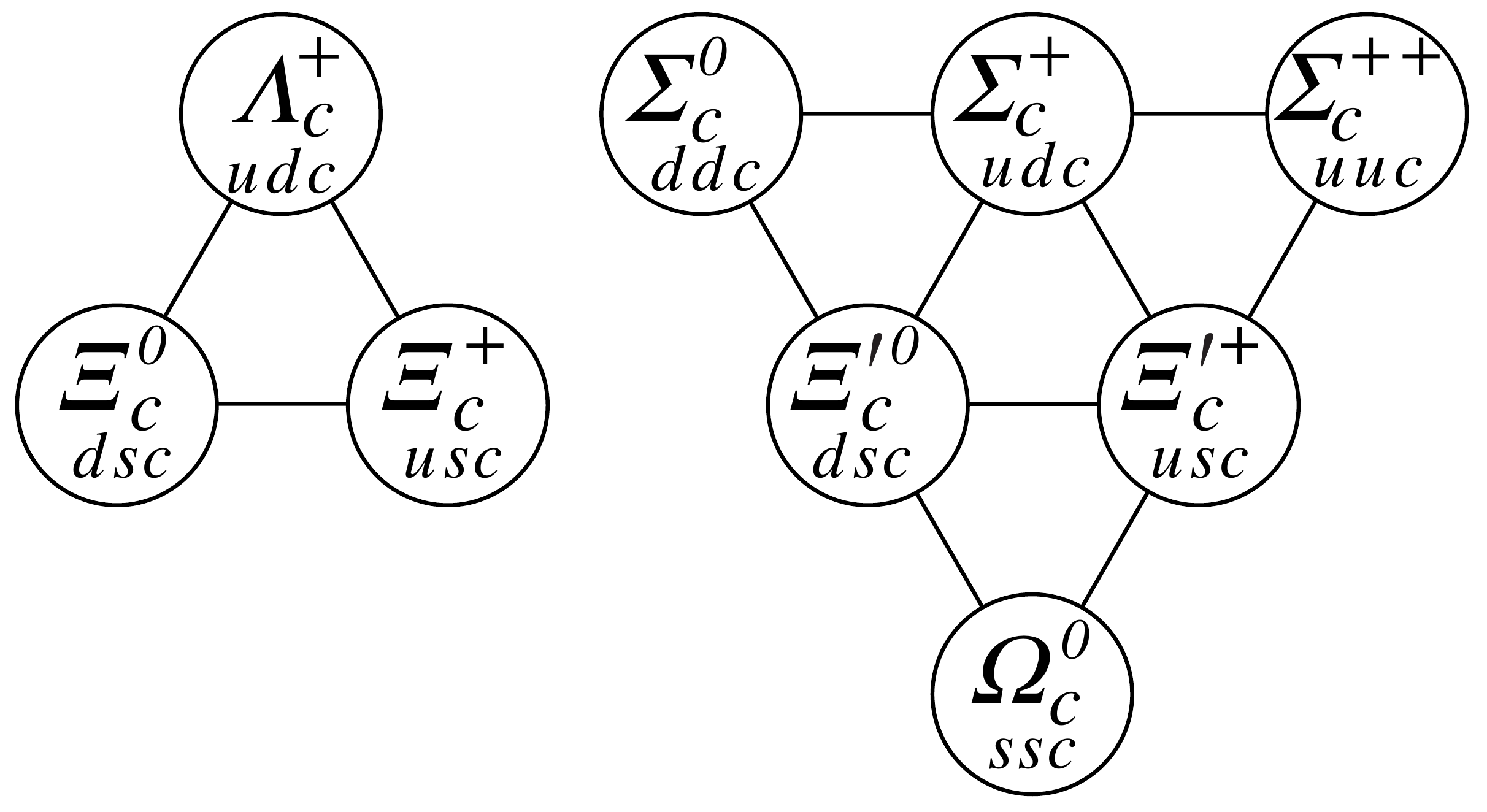}
\end{minipage} \hfill
\caption{The baryon multiplets constructed using the SU(4) group. The left diagram shows the spin-3/2 ${\bf 20}$-plet, the center diagram shows the spin-1/2 ${\bf 20^\prime}$-plet and in the right diagram we show the decomposition of the $c=1$ level of the spin-1/2 ${\bf 20^\prime}$-plet of the center diagram. All diagrams are taken from the PDG~\cite{Agashe:2014kda}.}  
\label{fig:spin12_32}
\end{figure}

The multiplet numerology is $4\otimes 4\otimes 4 = {\bf 20} \oplus {\bf 20_1^\prime} \oplus {\bf 20_2^\prime} \oplus {\bf \bar{4}}$. All the baryons in a given multiplet have the same spin and parity. Briefly, the ${\bf 20}$-plet consists of the spin-3/2 baryon states and can be further decomposed according to the charm content of the baryons into ${\bf 20} = {\bf 10} \oplus {\bf 6} \oplus {\bf 3} \oplus {\bf 1}$, where the ${\bf 10}$ is the standard $c=0$ decuplet and ${\bf 1}$ is the triply charm $\Omega_{ccc}^{++}$ singlet. The singly charmed baryon states belonging to the ${\bf 6}$ multiplet are symmetric under the interchange of $u$, $d$ and $s$ quarks, following the rule that the diquark $\left[(q_1)_a^T C\gamma_\mu (q_2)_b\right]$ is symmetric under interchanging $q_1\leftrightarrow q_2$. The doubly charmed ${\bf 3}$-plet consists of the isospin partners $\Xi^*_{cc}$ and the singlet $\Omega_{cc}^{*+}$. The ${\bf 20}$-plet is shown schematically in the left panel of~\fig{fig:spin12_32}. The corresponding interpolating fields of the spin-3/2 baryons are collected in~\tbl{spin32_tab} of Appendix A.

Similarly, the ${\bf 20^\prime}$-plet consists of the spin-1/2 baryons as shown schematically in the center panel of~\fig{fig:spin12_32}. It can be decomposed as ${\bf 20^\prime} = {\bf 8} \oplus {\bf 6} \oplus {\bf \bar{3}} \oplus {\bf 3}$. The ground level where $c=0$ comprises the well-known baryon octet, whereas the first level $c=1$ splits into two SU(3) multiplets, a ${\bf 6}$ and a ${\bf \bar{3}}$. The states of the ${\bf 6}$ are symmetric under interchanging $u$, $d$ and $s$ where the states of the ${\bf \bar{3}}$ are anti-symmetric. We show these states explicitly in the right panel of~\fig{fig:spin12_32}. We note that the diquark $\left[(q_1)_a^T C\gamma_5 (q_2)_b\right]$ appearing in the interpolating field of spin-1/2 baryons is anti-symmetric under interchanging $q_1\leftrightarrow q_2$. The top level consists of the ${\bf 3}$-plet with $c=2$. The interpolating fields of the spin-1/2 baryons are collected in~\tbl{spin12_tab} of Appendix A. The fully antisymmetric ${\bf \bar{4}}$-plet is not considered in this work.

In order to suppress excited state contamination, we apply Gaussian smearing to the quark fields at the source and sink~\cite{Gusken:1989qx,Alexandrou:1992ti}, given by $q^{\rm smear}(\vec{x},t) = \sum_y F(\vec{x},\vec{y};U(t))q(\vec{y},t)$, where $F(\vec{x},\vec{y};U(t)) = \left(\mathbb{1}+\alpha_G H\right)^{n_G}(\vec{x},\vec{y};U(t))$ is the gauge invariant smearing function and $H$ is the hopping term realized as a matrix in coordinate, color and spin space,
\be \label{eq:hopping_matrix}
H(\vec{x},\vec{y};U(t))=\sum_{j=1}^3 \left(U_j(\vec{x},t)\delta_{\vec{x}+a\hat{j},\vec{y}}+U^\dag_j(\vec{x}-a\hat{j},t)\delta_{\vec{x}-a\hat{j},\vec{y}} \right).
%H(\vec{x},\vec{y};U(t))=\sum_{\mu=1}^3 \left((1-\gamma_\mu)U_\mu(\vec{x},t)\delta_{\vec{x}+a\hat{\mu},\vec{y}}+(1+\gamma_\mu)U^\dag_\mu(\vec{x}-a\hat{\mu},t)\delta_{\vec{x}-a\hat{\mu},\vec{y}} \right).
\ee
The parameters $\alpha_{\rm G}$ and $n_{\rm G}$ of the Gaussian smearing used in this work are $\alpha_{\rm G} = 4.0$ and $n_{\rm G} = 50$.

In addition, we apply APE smearing to the spatial links that enter the hopping term. The parameters of the APE smearing we used are $\alpha_{\rm APE} = 0.5$ and $n_{\rm APE} = 50$.

The interpolating fields for the spin-3/2 baryons as defined in~\tbl{spin32_tab} have an overlap with spin-1/2 states. In order to isolate the desired spin-3/2 ground state, we incorporate a spin-3/2 projector in the definitions of the interpolating fields
\be 
\mathcal{J}_{B_{3/2}}^\mu = P^{\mu\nu}_{3/2} \mathcal{J}_{\nu B} \, .
\ee
For non-zero momentum, $P^{\mu\nu}_{3/2}$ is defined by~\cite{Benmerrouche:1989uc}
\be \label{eq:proj32}
P^{\mu\nu}_{3/2} = \delta^{\mu\nu} - \reci{3}\gamma^\mu \gamma^\nu - \reci{3p^2}\left(\not{p}\gamma^\mu p^\nu + p^\mu \gamma^\nu \not{p}  \right) \, .
\ee
The corresponding spin-1/2 projector is obtained by $P^{\mu\nu}_{1/2}=\delta^{\mu\nu} - P^{\mu\nu}_{3/2}$. In this work we study the mass spectrum of the baryons in the rest frame taking $\vec{p}=\vec{0}$, thus the last term of \eq{eq:proj32} will vanish. When the spin-3/2 and spin-1/2 projectors are applied to the interpolating field operators, the resulting two-point correlators for the spin-3/2 baryons acquire the form
\bea \label{eq:correlators_32_12}
C_{\frac{3}{2}} (t) &=& \frac{1}{3}\tr [C(t)] + \frac{1}{6} \sum_{i\ne j}^3 \gamma_i \gamma_j C_{ij}(t)\;, \nonumber\\
C_{\frac{1}{2}} (t) &=& \frac{1}{3}\tr [C(t)] - \frac{1}{3} \sum_{i\ne j}^3 \gamma_i \gamma_j C_{ij}(t)\;,
\eea 
where $\tr[C] = \sum_i C_{ii}$. It turns out that for some of the baryons we study, such as the $\Delta$, the inclusion of the spin-3/2 projector does not have a significant effect in the correlation function. However, we find that it is necessary for some others, such as the $\Xi^*$, in order to isolate the spin-3/2 ground state. Therefore, in order to ensure that we always measure the desired spin-3/2 ground state, we apply the spin-3/2 projector to all of the interpolating fields of \tbl{spin32_tab}. The reader interested in more details on the effects of these projectors on the baryon two-point functions and masses is referred to Ref.~\cite{Alexandrou:2014sha}.

%=============================================================
%=============================================================

\subsection{Determination of the lattice spacing}\label{subsection:alat}

In order to fix the lattice spacing for our physical ensemble, we used the physical nucleon mass as input. For this purpose we carried out a dedicated high statistics analysis of the nucleon mass with around 800,000 measurements, leading to an accurate determination of the lattice spacing.
The pion and nucleon mass in lattice units are 
\be\label{eq:pion_nucl_mass}
am_\pi = 0.06208(2)\;\;,\; am_N = 0.4436(11)\;,
\ee
yielding a ratio of  $m_N / m_\pi = 7.15(2)$, which differs by $2.9\%$ when compared to the physical value of $0.938/0.135 = 6.948$. If we were to 
assume that we are exactly at the physical point and use the physical value of the nucleon mass we would obtain $a=0.4436/0.938=0.473(1)$GeV~$^{-1}=0.0932(2)$~fm, yielding $m_\pi=0.1312(3)$~GeV i.e. 3\% smaller than physical.
Allowing to  be slightly away from the physical pion mass 
we can perform an interpolation to the physical point as follows: Observing that our previous results using $N_f=2$ and $N_f=2+1+1$ ensembles showed no detectable cut-off and volume effects nor we have seen any unquenching effects due to the strange and charm quarks in the sea, we make use of  the nucleon masses from 17 $N_f=2+1+1$ ensembles~\cite{Alexandrou:2014sha} in order to interpolate the nucleon mass of the physical ensemble.
Namely, we perform a combined fit to the $N_f=2$ physical ensemble and the 17 $N_f=2+1+1$ ensembles using the SU(2) chiral perturbation theory ($\chi$PT) 
well-established $\mathcal{O}(p^3)$ expression~\cite{Gasser:1987rb,Tiburzi:2008bk}
\be \label{eq:nucleon_p3}
m_N = m_N^{(0)} - 4c_1m_\pi^2 - \frac{3g_A^2}{32\pi f_\pi^2} m_\pi^3\;.
\ee
We collect the pion and nucleon masses that we used in the fit in~\tbl{Table:nucleon_masses} of Appendix~\ref{app:tables}. The three lattice spacings of the $N_f=2+1+1$ ensembles, the lattice spacing of the physical ensemble as well as the nucleon mass at the chiral limit, $m_N^0$, are treated as fit parameters. The value of $c_1$ is fixed such that the fit curve passes through the physical value of the nucleon mass at physical pion mass (physical point).
In order to estimate the  systematic error due to the chiral extrapolation, we also perform the fit using heavy baryon chiral perturbation theory (HB$\chi$PT) to $\mathcal{O}(p^4)$ in the small scale expansion (SSE) scheme~\cite{Procura:2006bj}. This form includes explicit $\Delta$ degrees of freedom by introducing as an additional parameter the $\Delta$-nucleon mass splitting, $\Delta\equiv m_\Delta - m_N$, taking $\mathcal{O}(\Delta /m_N) \sim \mathcal{O} (m_\pi /m_N)$. For completeness, we give the expression for the nucleon mass in the SSE scheme
\bea \label{eq:nucleon_p4}
	m_N &=& m_N^0-4c_1 m_\pi^2-\frac{3g_A^2}{32\pi f_\pi^2} m_\pi^3 -4E_1(\lambda)m_\pi^4 - \frac{3\left(g_A^2+3c_A^2 \right)}{64\pi^2f_\pi^2 m_N^0}m_\pi^4 - \frac{\left(3g_A^2+10c_A^2 \right)}{32\pi^2f_\pi^2 m_N^0}m_\pi^4 \log\left(\frac{m_\pi}{\lambda}\right) \nonumber\\	 
	&-& \frac{c_A^2}{3\pi^2f_\pi^2} \left(1+\frac{\Delta}{2m_N^0}\right)\left[\frac{\Delta}{4}m_\pi^2 + \left(\Delta^3-\frac{3}{2}m_\pi^2\Delta\right) \log\left(\frac{m_\pi}{2\Delta}\right) + \left(\Delta^2-m_\pi^2\right) R\left(m_\pi\right) \right]\quad,
\eea   
where $R\left(m_\pi\right)= - \sqrt{m_\pi^2 - \Delta^2}\cos^{-1}\left(\frac{\Delta}{m_\pi}\right)$ for $m_\pi > \Delta$ and $R\left(m_\pi\right)= \sqrt{\Delta^2 - m_\pi^2} \log\left( \frac{\Delta}{m_\pi}+\sqrt{\frac{\Delta^2}{m_\pi^2}-1} \right)$ for $m_\pi < \Delta$. We take the cut-off scale $\lambda=1$ GeV, $c_A=1.127$~\cite{Procura:2006bj} and treat the counter-term $E_1$ as an additional fit parameter. The physical values of $f_\pi$ and $g_A$ are used in both fits, namely $f_\pi=0.092419(7)(25)$ GeV and $g_A=1.2695(29)$. We take the difference between the results of the $\mathcal{O}(p^3)$ and $\mathcal{O}(p^4)$ fits as an estimate of the uncertainty due to the chiral extrapolation. The final value of the lattice spacing for the physical ensemble is
\be\label{eq:lat_spacing}
a_{N_f=2} = 0.0938(3)(2) \;  \rm{fm}\;,
\ee
where the error in the first parenthesis is the statistical error and the systematic error is given in the second parenthesis. This value is in agreement with the value extracted assuming the simulation is exactly at the physical point, which demonstrates that any deviation form the physical point is within the accuracy of the results.
From our lattice values of~\eq{eq:pion_nucl_mass} and using~\eq{eq:lat_spacing} we find that the pion mass in physical units is $m_\pi = 0.1305(4)$~GeV, which is about $3.3\%$ lower than the average physical pion mass, and the corresponding nucleon mass is $m_N = 0.9321(36)$~GeV, less than $1\%$ lower from the physical nucleon mass, which explains the agreement between the two determinations.

\begin{figure}[!ht]\vspace*{-0.2cm}
\begin{minipage}[t]{8.5cm}
\includegraphics[width=\linewidth]{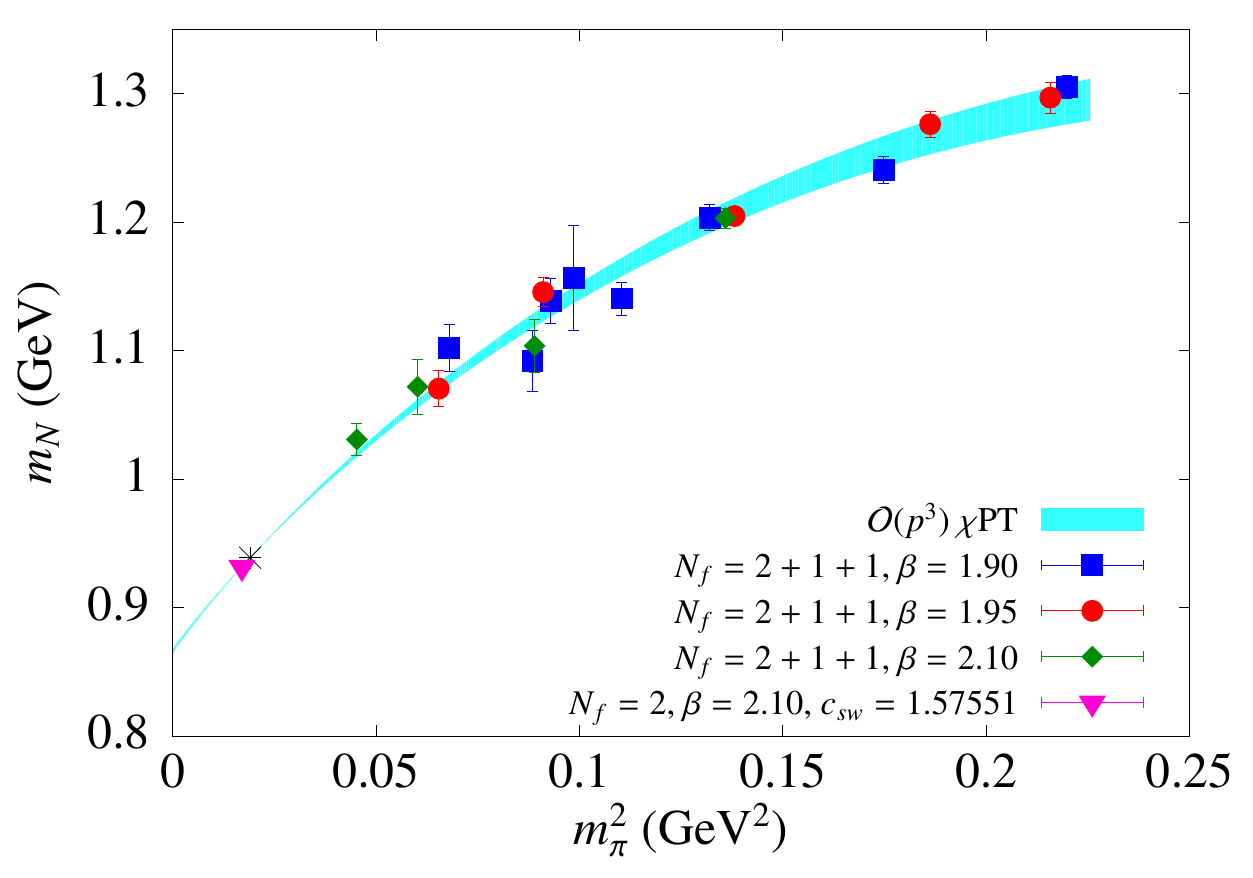}
\end{minipage} \hfill
\begin{minipage}[t]{8.5cm}
\includegraphics[width=0.995\linewidth]{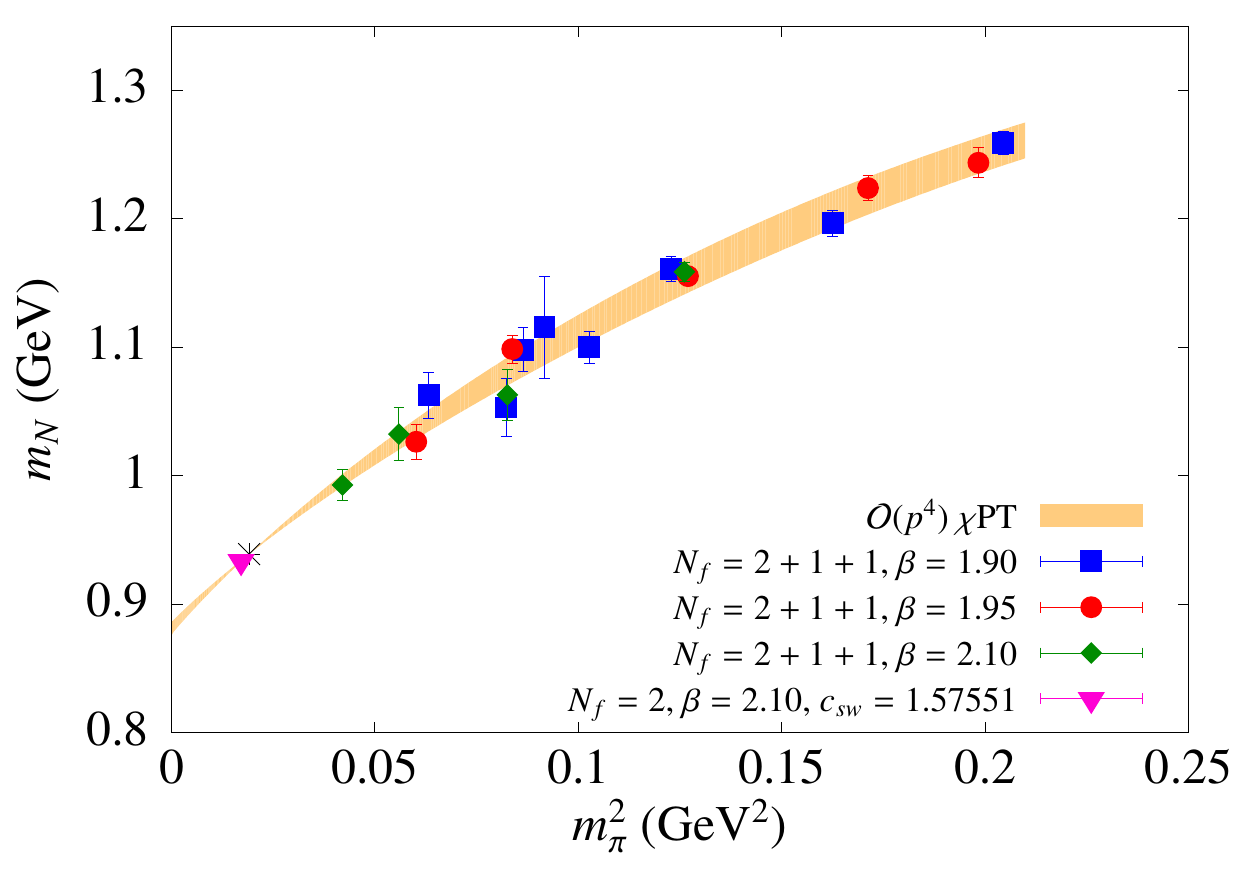}
\end{minipage} \hfill
\caption{Nucleon masses for the $N_f=2+1+1$ ensembles at $\beta=1.90$ (blue squares), $\beta=1.95$ (red circles) and $\beta=2.10$ (green diamonds) as well as for the physical ensemble (magenta triangle). The lowest order $\mathcal{O}(p^3)$ fit is shown on the left plot with the blue error band, whereas the $\mathcal{O}(p^4)$ fit is shown on the right plot and with the brown error band. The physical nucleon mass is denoted with the asterisk.}
\label{Fig:nucleon_mass}
\end{figure}

In \fig{Fig:nucleon_mass} we show the fits of the nucleon mass to the $\mathcal{O}(p^3)$ and $\mathcal{O}(p^4)$ expressions of Eqs.(\ref{eq:nucleon_p3}) and (\ref{eq:nucleon_p4}), respectively. The error band and the errors on the fit parameters are obtained from super-jackknife analysis~\cite{Bratt:2010jn}. As mentioned above, cut-off effects were investigated in Ref.~\cite{Alexandrou:2014sha} and were found to be negligible for the nucleon mass. This is corroborated by fitting the data for each $\beta$ of the $N_f=2+1+1$ ensembles separately to extract the lattice spacings. We find that the values are in agreement with those from the combined fit. We note that our lattice results exhibit a curvature, which supports the presence of the $m_\pi^3$-term. We remark here that by including the nucleon mass from the physical ensemble in the fit, the lattice spacings of the $N_f=2+1+1$ ensembles as well as the rest of the fitting parameters remain completely unchanged. In addition, if we fit using $N_f=2$ ensembles by ETMC~\cite{Alexandrou:2010hf,Alexandrou:2012xk} instead of the $N_f=2+1+1$ ensembles we obtain the same result as in~\eq{eq:lat_spacing} for the physical ensemble, and the lattice spacings for the $N_f=2$ ensembles have the same values, as if we fitted without the physical ensemble. These are indications that the interpolation carried out is very robust. The fit parameters for the two fits including the $\chi^2$/d.o.f. are given in \tbl{Table:nucleon_fitparams}. For completeness, we give the lattice spacings for the $N_f=2+1+1$ ensembles
\bea \label{eq:nf211_alat}
a_{\beta=1.90} &=& 0.0936(13)(35)\; \rm{fm}\;, \nonumber\\
a_{\beta=1.95} &=& 0.0823(10)(35)\; \rm{fm}\;, \nonumber\\
a_{\beta=2.10} &=& 0.0646(7)(25) \;  \rm{fm}\;,
\eea
where the error in the first parenthesis is statistical and in the second parenthesis the systematic due to the chiral extrapolation, as explained above.
\begin{table}[h]
\begin{center}
\renewcommand{\arraystretch}{1.2}
\renewcommand{\tabcolsep}{7.5pt}
\begin{tabular}{l|ccccc}
\hline\hline
 & $m_N^0$  &  $-4c_1 (\rm{GeV}^{-1})$ & $E_1(\lambda)$ (GeV$^{-3}$) & $\sigma_{\pi N}$ (MeV) & $\chi^2/{\rm d.o.f}$   \\				
\hline
$\mathcal{O}(p^3)$ HB$\chi$PT &  0.8667(15) & 4.5735  &               &  64.9(1.5) &  1.5779 \\
$\mathcal{O}(p^4)$ SSE        &  0.8813(47) & 3.7282  & -2.5858(2480) &  51.7(4.3) &  1.0880 \\
\hline\hline
\end{tabular}
\end{center}
\caption{Fit parameters $m_N^0$ in GeV and $E_1(\lambda)$ in GeV$^{-3}$ from $\mathcal{O}(p^3)$ HB$\chi$PT and $\mathcal{O}(p^4)$ SSE, as well as the fixed value of $-4c_1$. Also included is the value of the  $\sigma$-term for each fit. }
\label{Table:nucleon_fitparams}
\end{table}

Finally, we note that the value of~\eq{eq:lat_spacing} is fully consistent with the one determined from gluonic quantities, from $r_0$, and the ones related to the action density renormalised through the gradient flow. It is, however, larger by about 2\% as compared to the one extracted using $f_\pi$ and $f_K$~\cite{Abdel-Rehim:2015pwa}. We will use the lattice spacing given in~\eq{eq:lat_spacing} to convert to physical units all the quantities studied in this work.

Having determined the parameters of the chiral fit we can compute the nucleon $\sigma_{\pi N}$-term  by evaluating $m_\pi^2 \partial m_N/\partial m_\pi^2$ where we have taken the leading order relation $m_\pi^2 \sim \mu_l$.  Using \eq{eq:nucleon_p3} we find $\sigma_{\pi N} = 64.9\pm1.5$ MeV. Performing the same calculation using the $\mathcal{O}(p^4)$ expression we obtain a lower value of $\sigma_{\pi N} = 51.7\pm4.3$ MeV showing the sensitivity of this quantity to the chiral extrapolation. As with the fit parameters, the values of $\sigma_{\pi N}$-term are unchanged by including the nucleon mass at the physical ensemble in the fits~\cite{Alexandrou:2014sha}. We  note that these values are larger as compared to direct evaluations of this quantity by a number of lattice QCD groups including one performed using this ensemble~\cite{Abdel-Rehim:2016won}, where a value of $\sigma_{\pi N}=37.2(2.6)(^{4.7}_{2.9})$~MeV was obtained. Given the large variation when using the two different chiral expansions, the evaluation of $\sigma_{\pi N}$ from the slope of the fit receives a large systematic error of 13.2~MeV, giving   a value of $\sigma_{\pi N} = 64.9\pm1.5\pm 13.2$~MeV, which brings the disagreement with  the direct determination to one standard deviation.

%=============================================================
%=============================================================

\subsection{Tuning of the bare strange and charm quark masses}

\begin{figure}[!ht]\vspace*{-0.2cm}
\center
\begin{minipage}{8.5cm}
{\includegraphics[width=0.9\textwidth]{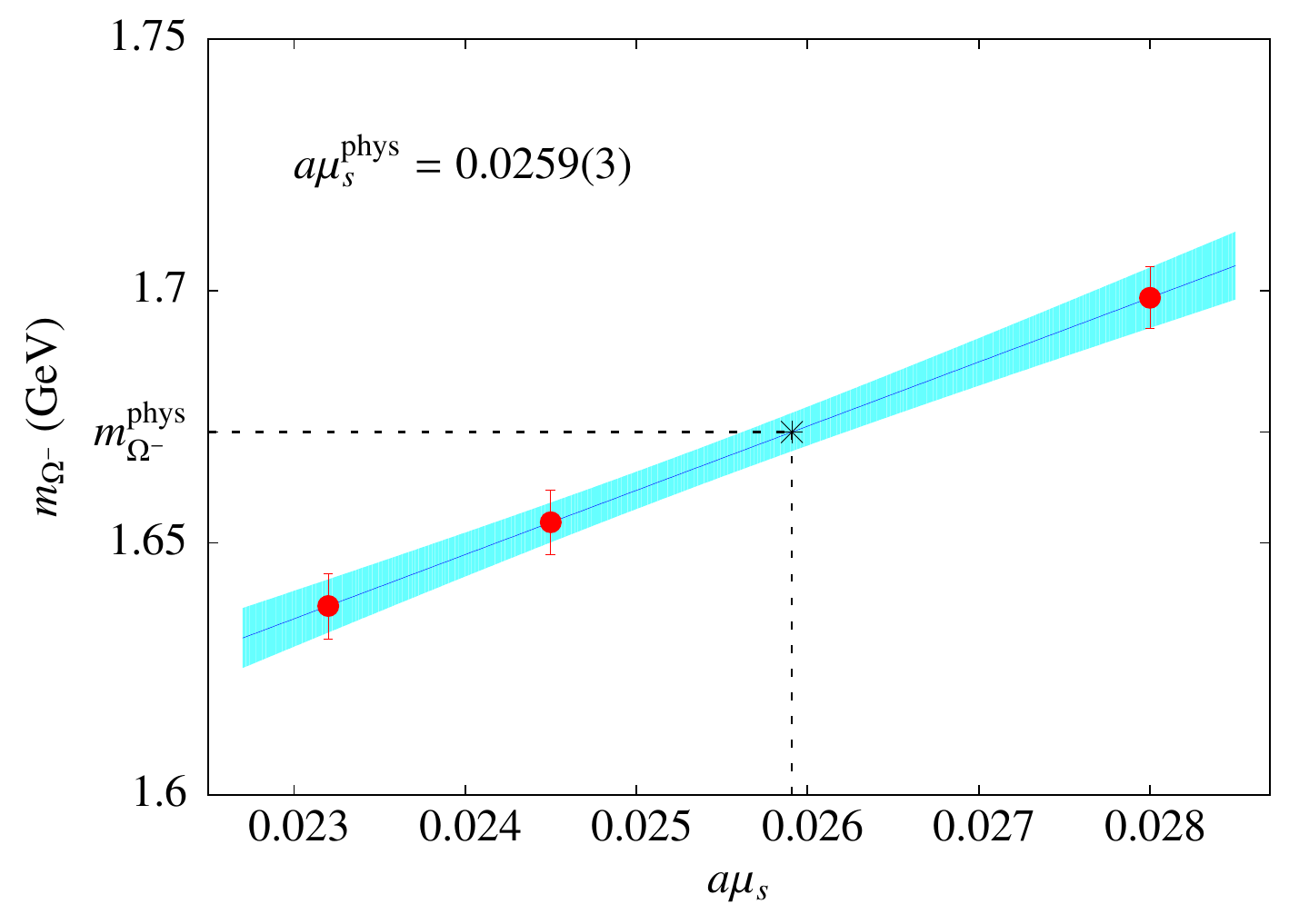}}
\end{minipage}\hfill
\begin{minipage}{8.5cm}
{\includegraphics[width=0.905\textwidth]{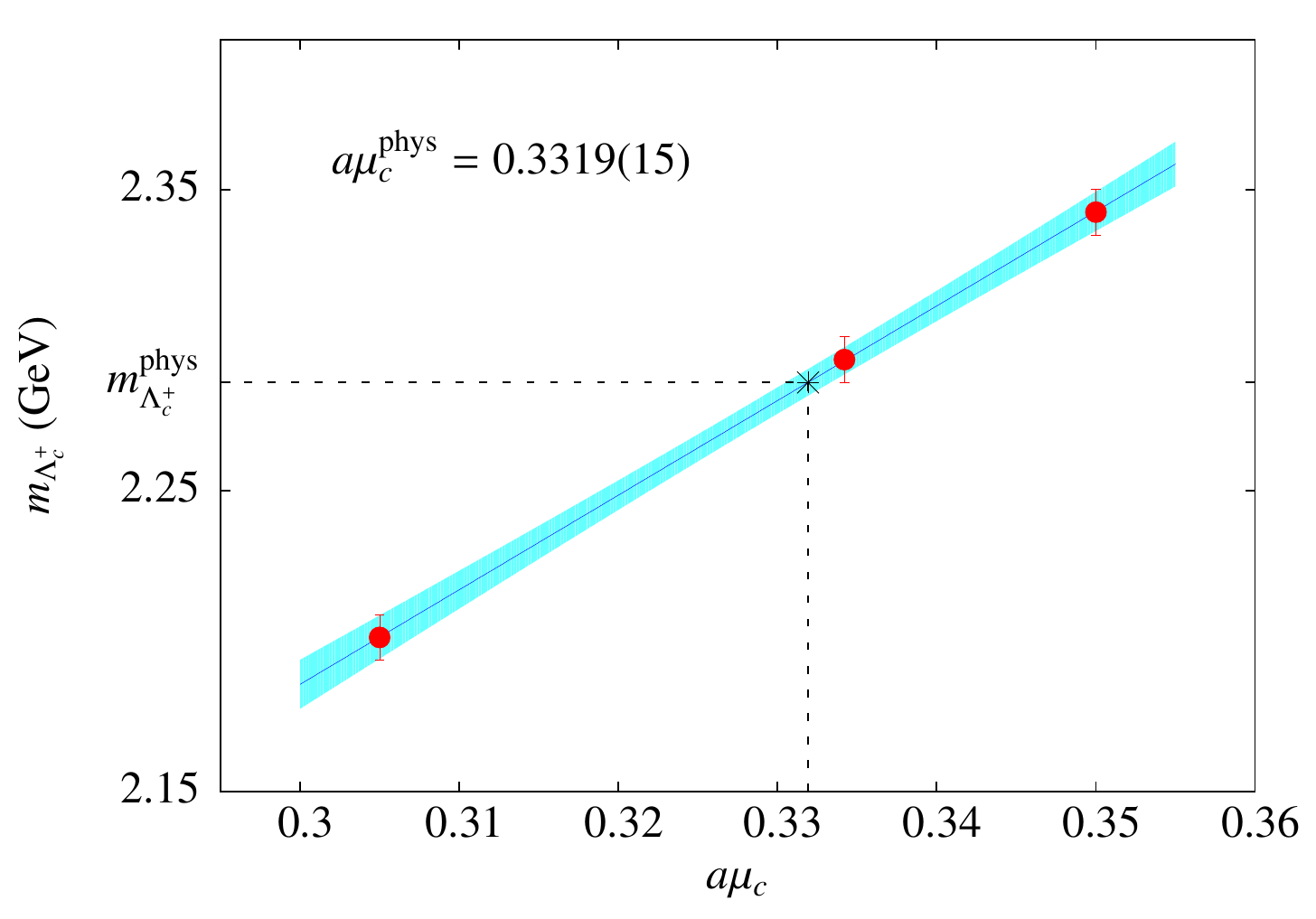}}
\end{minipage}
\caption{Tuning of the bare strange and charm quark masses with the experimental values of the $\Omega^-$ (left) and $\Lambda_c^+$ (right) masses respectively.}
\label{Fig:omega_lambda_ms_mc}
\end{figure}

In order to determine the bare strange and charm quark masses, we perform a tuning using the physical mass of the $\Omega^-$(1.672) baryon and the $\Lambda_c^+$(2.286) baryon, respectively, as input.  Our strategy is to calculate the $\Omega^-$ and $\Lambda_c^+$ masses at various trial values of $a\mu_s$ and $a\mu_c$ and then match directly with the physical $\Omega^-$ and $\Lambda_c^+$ mass, respectively, assuming small cut-off and finite volume effects. This procedure determines the tuned values of $a\mu_s$ and $a\mu_c$. In \fig{Fig:omega_lambda_ms_mc} we show the matching of the strange and charm quark masses with the physical $\Omega^-$ and $\Lambda_c^+$ masses, respectively. The values of $a\mu_s$ and $a\mu_c$ used for the tuning, along with the respective $\Omega^-$ and $\Lambda_c^+$ masses are listed in \tbl{Table:quark_masses_tuning}. 
An analysis using the same ensemble as the one we are using here yielded $a\mu_s = 0.0249(1)$ and $a\mu_c = 0.3075(15)$ from interpolation of the meson mass ratios $m_K/m_\pi$ and $m_D/m_\pi$~\cite{Abdel-Rehim:2015pwa}, showing an agreement within 4\% and 7\%, respectively, when compared to our results. This is very satisfactory given that systematic errors are not included. Since we interested in the baryon sector we use the tuned quark mass values determined from using baryonic observables. The tuned values we find for the bare heavy quark masses are 
\bea \label{eq:strange_charm_values}
a\mu_s &=& 0.0259(3)  \nonumber\\
a\mu_c &=& 0.3319(15) \;, 
\eea
where the error is the statistical, obtained from the fit band. 
From these values we find $\mu_s/\mu_l = 28.8(3)$ and $\mu_c/\mu_s = 12.8(2)$. Our analysis using the meson mass ratios $m_K/m_\pi$ and $m_D/m_\pi$ for the same ensemble as the one we are using here are $\mu_s/\mu_l = 27.7(1)$ and $\mu_c/\mu_s = 12.3(1)$~\cite{Abdel-Rehim:2015pwa}. These ratios are about a standard deviation different from the ones we find in this work,  indicating that systematic errors on these ratios from using different quantities to fix the quark masses are small and comparable with the statistical ones. 

The renormalization constant $Z_P$ is determined for this ensemble non-perturbatively. We find $Z_P = 0.501(8)(26)(12)$ in the $\overline{\rm MS}$ at 2 GeV~\cite{Abdel-Rehim:2015owa}, where the first error is statistical, the second is a systematic error stemming from the extrapolation to $(ap)^2 = 0$ and the perturbative subtraction of leading lattice artefacts, and the third from the conversion of RI$^\prime$-MOM to $\overline{\rm MS}$ at 2 GeV. Using this value of $Z_P$ and the lattice spacing of~\eq{eq:lat_spacing}, the renormalized strange and charm quark masses are  

\be m_s^R = \mu_s/Z_P=108.6(2.2)(5.7)(2.6)~{\rm MeV}\,\, {\rm and}
\,\, m_c^R = \mu_c/Z_P=1.39(2)(7)(3)~{\rm GeV}\;,
\label{renormalized mass}
\ee
where the first error is statistical, the second the combined systematic error from the determination of $Z_P$ and the lattice spacing of~\eq{eq:lat_spacing} and the third from the conversion of RI$^\prime$-MOM to $\overline{\rm MS}$ at 2 GeV. The corresponding renormalized masses determined from meson mass ratios $m_K/m_\pi$ and $m_D/m_\pi$ for the same ensemble are $m_s^R=107(2)(6)(3)$~MeV and $m_c^R=1.33(3)(7)(3)$~GeV in the $\overline{\rm MS}$ at 2 GeV~\cite{Abdel-Rehim:2015pwa} with the errors being determined in the same manner as in~\eq{renormalized mass}. These renormalized strange and charm quark masses are in agreement with the values given in~\eq{renormalized mass}. A more complete analysis, including systematic errors due to lattice artefacts will follow in the future.

It is interesting to compare our values of the strange and charm quark masses with the ones given by the FLAG group. The $N_f=2$ FLAG ratios are $m_s/m_l=27.3(9)$ and $m_c/m_s=11.74(35)$~\cite{Aoki:2016frl}. The FLAG values are continuum extrapolated and corrected for finite volume effects. The fact that our values are within one standard deviation for the $m_s/m_l$ and two standard deviations for the $m_c/m_s$ is very satisfactory. 
Furthermore, the $N_f=2$ $m^R_s$ and $m^R_c$ values obtained by the FLAG are~\cite{Aoki:2016frl}
\be m_s^R (2{\rm GeV})= 101(3)~{\rm MeV}\,\, {\rm and}
\,\, m_c^R(2 {\rm GeV}) = 1.14 (4)~{\rm GeV}.
\label{FLAG values}
\ee
in the $\overline{\rm MS}$, where the value for $m_c^R$ resulted from an analysis  using twisted mass ensembles from the meson sector~\cite{Blossier:2010cr}.  The strange renormalized mass quoted by FLAG is consistent with our value determined from the $\Omega^-$ at this fixed lattice spacing. The  renormalized charm quark mass is smaller by two standard deviations, which is rather satisfactory given that our value is obtained for one ensemble with no evaluation of cut-off effects. 

\begin{table}[h]
\begin{center}
\renewcommand{\arraystretch}{1.25}
\renewcommand{\tabcolsep}{5pt}
\begin{tabular}{l|ll||l|ll}
\hline\hline
$\; a\mu_s$ & $\quad am_{\Omega^-}$  & $ m_{\Omega^-}$ (GeV) & $\; a\mu_c$  & $\quad am_{\Lambda_c^+}$  & $ m_{\Lambda_c^+}$ (GeV)  \\
\hline
0.0232 & 0.7793(31)   & 1.6375(65)   & 0.3050 &  1.0475(36)  &  2.2012(75)  \\
0.0245 & 0.7872(30)   & 1.6541(64)   & 0.3342 &  1.0915(36)  &  2.2936(76)  \\
0.0280 & 0.8084(29)   & 1.6987(61)   & 0.3500 &  1.1149(37)  &  2.3427(77)  \\
\hline\hline
\end{tabular}
\caption{Masses of the $\Omega^-$ and $\Lambda_c^+$ baryons  at the trial values of $a\mu_s$ and $a\mu_c$ in lattice and physical units with the associated statistical error. The lattice spacing value of~\eq{eq:lat_spacing} was used for converting to physical units.}
\label{Table:quark_masses_tuning}
\end{center}
\vspace*{-.0cm}
\end{table}

We note here that our correlation functions were produced with $am_s = 0.0264$ and $am_c = 0.3348$. In order to correct for this small difference, we interpolate our results to the tuned values of~\eq{eq:strange_charm_values} using our results at the three different $am_s$ and $am_c$ listed in~\tbl{Table:quark_masses_tuning}.

%=============================================================
%=============================================================

\section{Lattice Results}\label{sec:latresults}

\subsection{Isospin symmetry breaking}

The breaking of the isospin symmetry is a feature of the lattice twisted mass fermion action due to the presence of $\tau^3$ acting in flavor space. Isospin breaking effects are of the order $\mathcal{O}(a^2)$ and in general they are detectable as mass splittings between hadrons belonging to the same isospin multiplets. Possible isospin splitting effects should vanish in the continuum limit. There is still an exact symmetry of the twisted mass action, namely parity combined with an interchange of u- and d-quarks, according to which the proton and the neutron are degenerate, as are the $\Delta^{++}$, $\Delta^-$ and the $\Delta^+$, $\Delta^0$ baryons. However, there could be a mass difference between, e.g. the $\Delta^{++}$ and the $\Delta^+$ baryons. Therefore, we average over the masses of the proton and the neutron, as well as the $\Delta^{++}$, $\Delta^-$ and $\Delta^+$, $\Delta^0$. In the latter case, we take the difference between the two averages to study isospin splitting effects. We extend the isospin breaking study for all isospin multiplets of the forty baryons we analyze in this work. In all figures concerning isospin splitting, we additionally show the corresponding splitting for the $N_f=2+1+1$ ensembles, analyzed in a previous work~\cite{Alexandrou:2014sha} for comparison.

\begin{figure}[!ht]\vspace*{-0.2cm}
\center
\begin{minipage}{8.5cm}
{\includegraphics[width=0.9\textwidth]{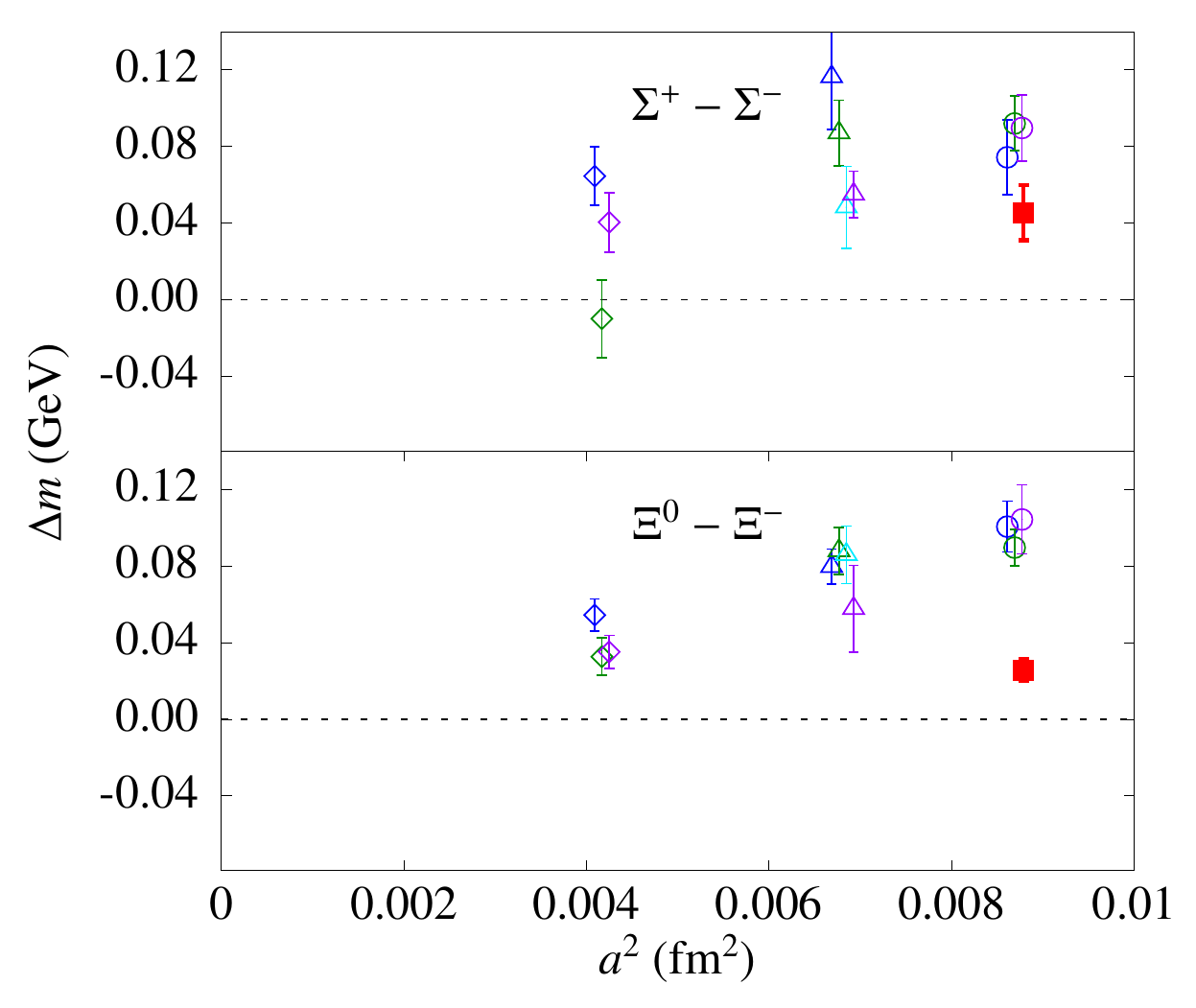}}
\end{minipage}\hfill
\begin{minipage}{8.5cm}
{\includegraphics[width=0.905\textwidth]{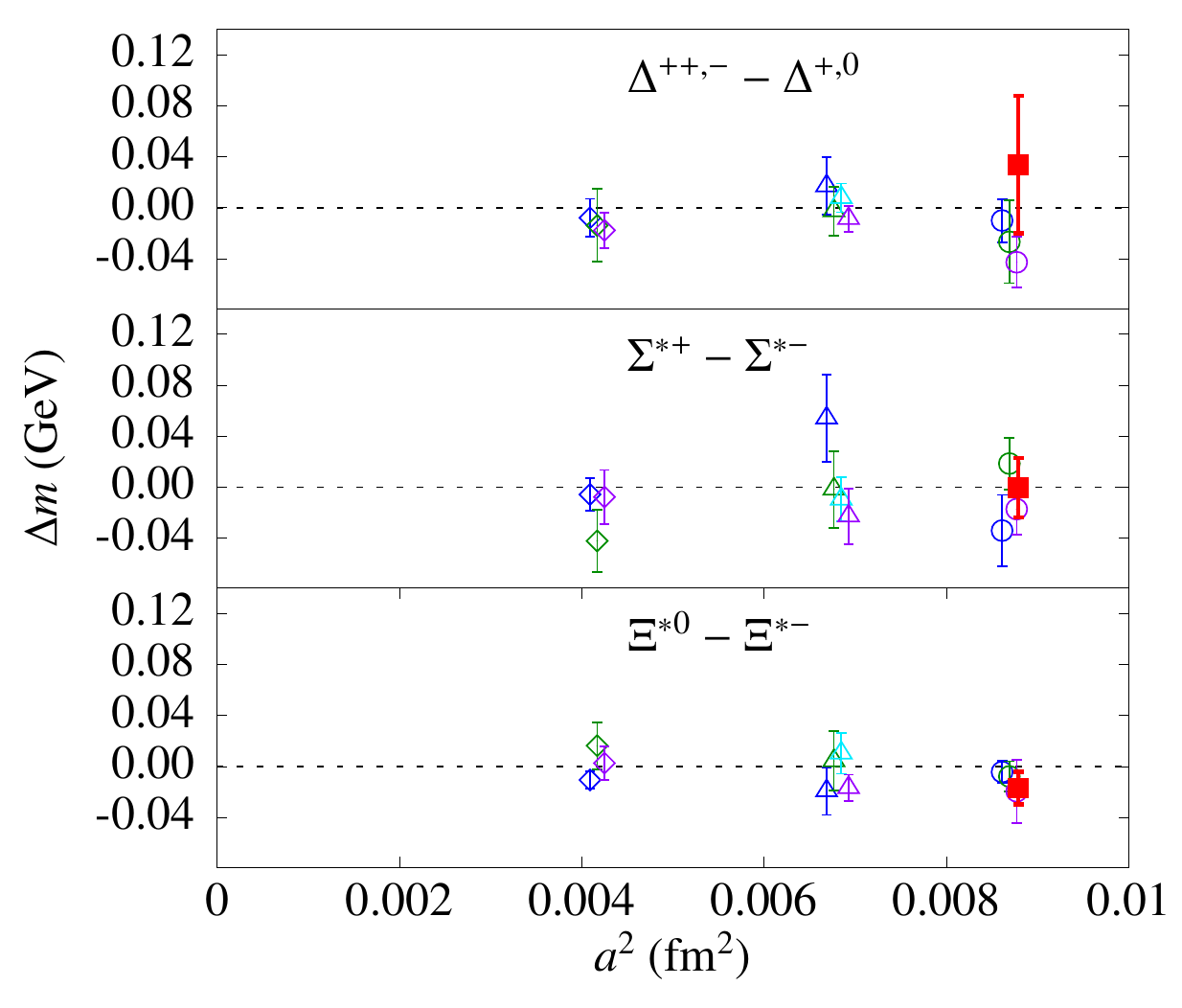}}
\end{minipage}
\caption{Mass difference for the octet (left) and decuplet (right) baryons, as a function of the lattice spacing squared. Results from this work at the physical ensemble are shown with the red filled square. With open symbols we show the results from Ref.~\cite{Alexandrou:2014sha} using $N_f=2+1+1$ ensembles at $a=0.094$ (circles), $a=0.082$ (triangles) and $a=0.065$ (diamonds), with the different colours denoting the various pion masses at each lattice spacing (blue for lightest pion mass, purple for heaviest pion mass). Some of the points of the $N_f=2+1+1$ results at each lattice spacing have been slightly shifted to the left and right for clarity.}
\label{Fig:8_10_isospin}
\end{figure}

We start this analysis by showing the mass difference for the octet and decuplet isospin multiplets, shown in~\fig{Fig:8_10_isospin}. In the octet case there are small mass splittings in the $\Sigma$ and $\Xi$ baryon multiplets, which amount to about $3\%$ of the mass of the baryons at the isospin limit. This splitting is taken as a systematic error in our final results for the $\Sigma$ and $\Xi$ baryons. It is also notable that the breaking is more than twice  smaller when compared to the corresponding ones obtained using  the $N_f=2+1+1$ ensembles at similar value of the lattice spacing, which confirms that combining Wilson twisted mass fermions at maximal twist and the clover term reduces cut-off effects related to isospin symmetry breaking. Regarding the decuplet, the mass difference in the $\Delta$, $\Sigma^*$ and $\Xi^*$ isospin multiplets is consistent with zero within our statistical accuracy, indicating that isospin splitting effects are minimal in this case.

\begin{figure}[!ht]\vspace*{-0.2cm}
\center
\begin{minipage}{8.5cm}
{\includegraphics[width=0.9\textwidth]{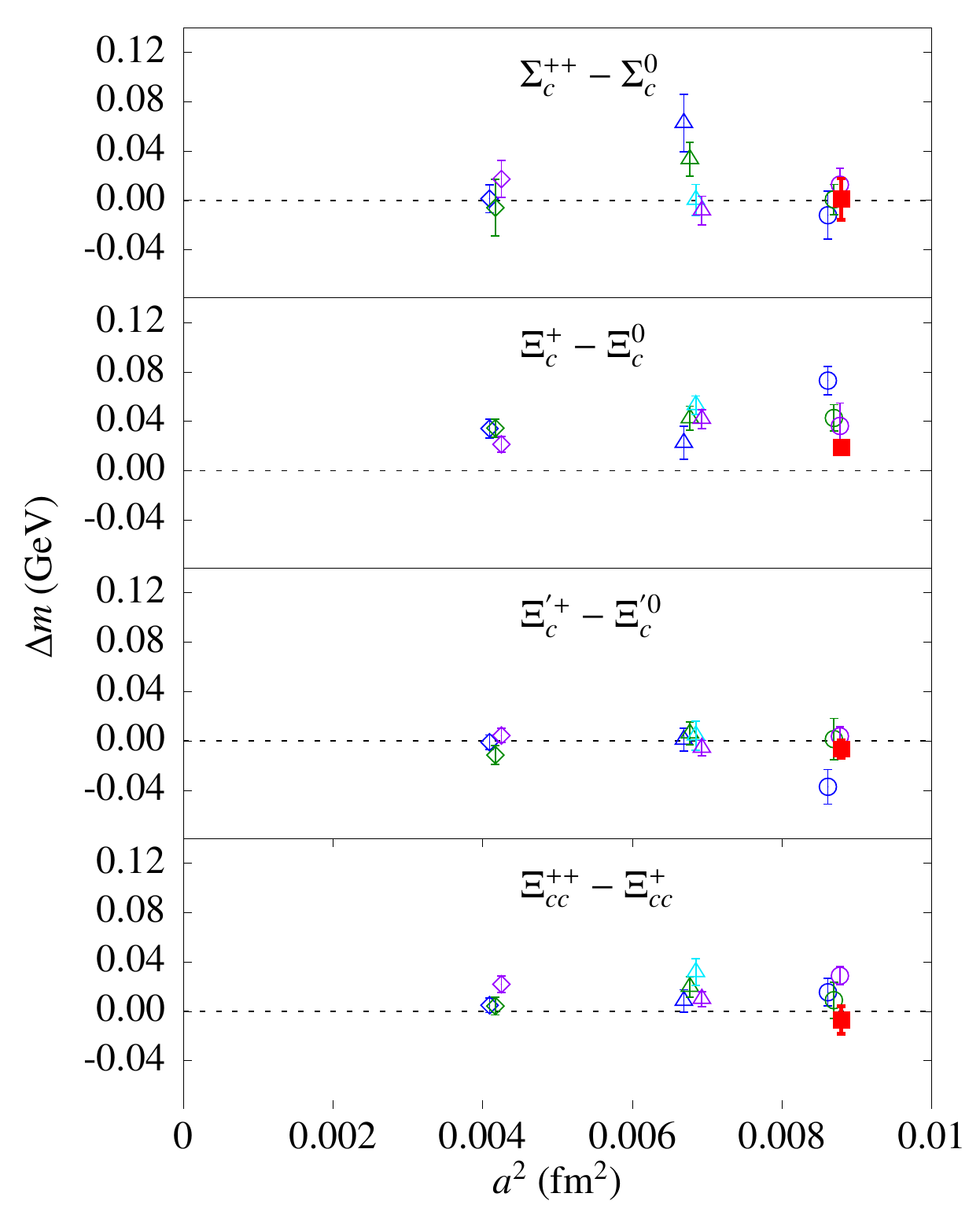}}
\end{minipage}\hfill
\begin{minipage}{8.5cm}
{\includegraphics[width=0.905\textwidth]{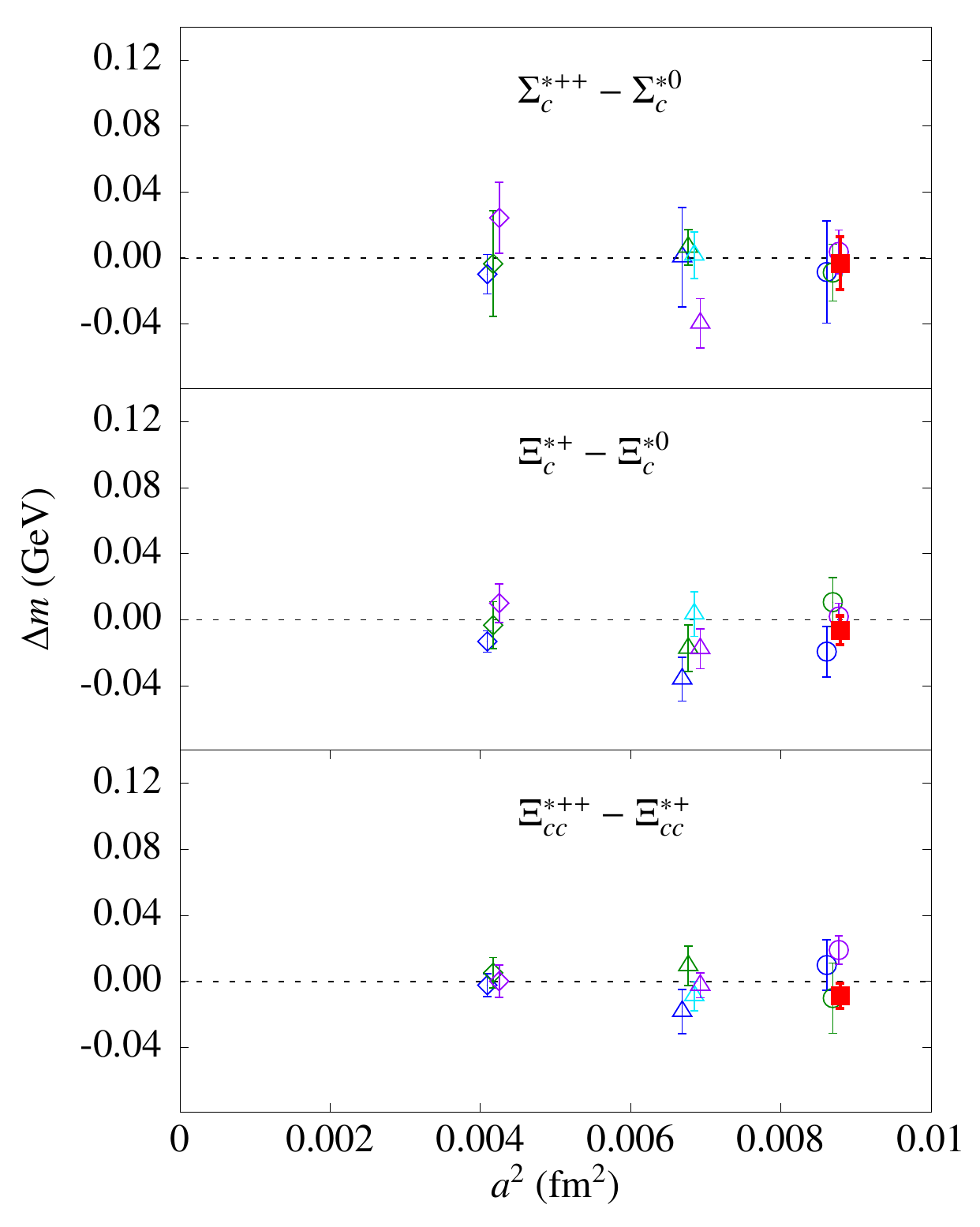}}
\end{minipage}
\caption{Mass difference for the charm spin-1/2 (left) and spin-3/2 (right) baryons, as a function of the lattice spacing squared. The notation is the same as in~\fig{Fig:8_10_isospin}.}
\label{Fig:charm_isospin}
\end{figure}

In the charm sector, we show the mass difference of the spin-1/2 $\Sigma_c$, $\Xi_c$, $\Xi_c^\prime$ and $\Xi_{cc}$ multiplets in the left panel of~\fig{Fig:charm_isospin}. As can be seen, the mass splitting is consistent with zero for all states except $\Xi_c$, where a mere $1\%$ splitting is observed. As with the decuplet, the charm spin-3/2 multiplets $\Sigma_c^*$, $\Xi_c^*$ and $\Xi_{cc}^*$ display  zero mass splitting, as it is shown in the right panel of~\fig{Fig:charm_isospin}.

These observations lead to the conclusion that the isospin symmetry breaking for our physical ensemble is either consistent with zero or   smaller than 3\%. In what follows we will average over the masses of the various isospin multiplets to obtain the final values of their mass.

%=============================================================
%=============================================================

\subsection{Final results and comparison}\label{sec:comparison}

In this section we present our final results for the low-lying baryon masses studied in this work using our physical ensemble. We use the lattice spacing of~\eq{eq:lat_spacing} to convert to physical units. We give the final results in~\tbl{Table:phys_masses}, where in  the first parenthesis we give the statistical error. We estimate a systematic error due to the tuning of the heavy quark masses, shown in the second parenthesis, by interpolating our lattice results to the larger and smaller values of the strange and charm quark masses allowed by the errors of Eq.~\ref{eq:strange_charm_values}. For the $\Sigma$, $\Xi$ and $\Xi_c$ baryons we additionally take into account the non-zero isospin splitting effects by including a systematic error as the mass difference between the associated isospin partners in these multiplets, shown in the third parenthesis.

\begin{table}[!ht]
\begin{center}
\renewcommand{\arraystretch}{1.4}
\renewcommand{\tabcolsep}{7.5pt}
\begin{tabular}{cccccc}
\multicolumn{6}{c}{Octet and decuplet baryons}\\
\hline
$\Lambda$ (1.116)  & $\Sigma$ (1.193)   &   $\Xi$ (1.318)  &  $\Delta$ (1.232)  &   $\Sigma^*$ (1.384)   &  $\Xi^*$ (1.530) \\
%\hline
1.108(8)(2) & 1.193(13)(3)(45) &  1.305(8)(7)(26)  & 1.225(59) &  1.416(23)(15) &  1.525(17)(15) \\
\hline\hline \\ %[0.02cm]
\multicolumn{6}{c}{Spin-1/2 charm baryons}\\
\hline
%\end{tabular} \\% [0.1cm]
%
%\begin{tabular}{cccccc}
%\hline\hline
$\Sigma_c$ (2.453)    &  $\Xi_c$  (2.470)   &  $\Xi_{c}^\prime$(2.575)   &  $\Omega_c^0$ (2.695)  &  $\Xi_{cc}$ (3.519)  &  $\Omega_{cc}^{+}$    \\
%\hline
2.468(18)(10) & 2.465(7)(10)(19) & 2.579(10)(3) & 2.685(7)(12) & 3.606(11)(8) & 3.711(5)(30)     	 \\
\hline\hline \\
%\end{tabular} \\% [0.1cm]
%
%\begin{tabular}{cccccc}
%\hline\hline
\multicolumn{6}{c}{Spin-3/2 charm baryons}\\
\hline
$\Sigma_c^*$ (2.517)  &  $\Xi_c^*$ (2.645)  &  $\Omega_c^{*0}$ (2.765)   &  $\Xi_{cc}^*$          &  $\Omega_{cc}^{*+}$  &  $\Omega_{ccc}^{++}$  \\
%\hline
2.539(18)(22) & 2.641(13)(8) & 2.746(7)(28) & 3.682(10)(26) & 3.770(6)(30) & 4.746(4)(32)     	 \\
\hline\hline
\end{tabular} \\ [0.1cm]
\end{center}
\caption{The values of the masses of the baryons considered in this work after converting to physical units and averaging over the various multiplets, with the associated statistical error in the first parenthesis and the systematic error due to the tuning in the second parenthesis. For the $\Sigma$, $\Xi$ and $\Xi_c$ baryons the systematic error due to the isospin splitting is shown in the third parenthesis. The experimental mass for each baryon~\cite{Agashe:2014kda}, wherever exists, is shown in parenthesis next to its symbol. The mass of the nucleon, $\Omega^-$ and $\Lambda_c^+$ are omitted, since they are used as input to the calculations.}
\label{Table:phys_masses}
\end{table}

We compare the results given in Table~\ref{Table:phys_masses} with a number of other lattice QCD calculations using different discretization schemes. We also include our previous results obtained using $N_f=2+1+1$ twisted mass gauge configurations. The results from all other lattice calculations referred to from now on are extrapolated to the physical point unless otherwise specified. We state explicitly, which calculations have also taken the continuum limit.

\begin{figure}[!ht]
\center
{\includegraphics[width=0.6\textwidth]{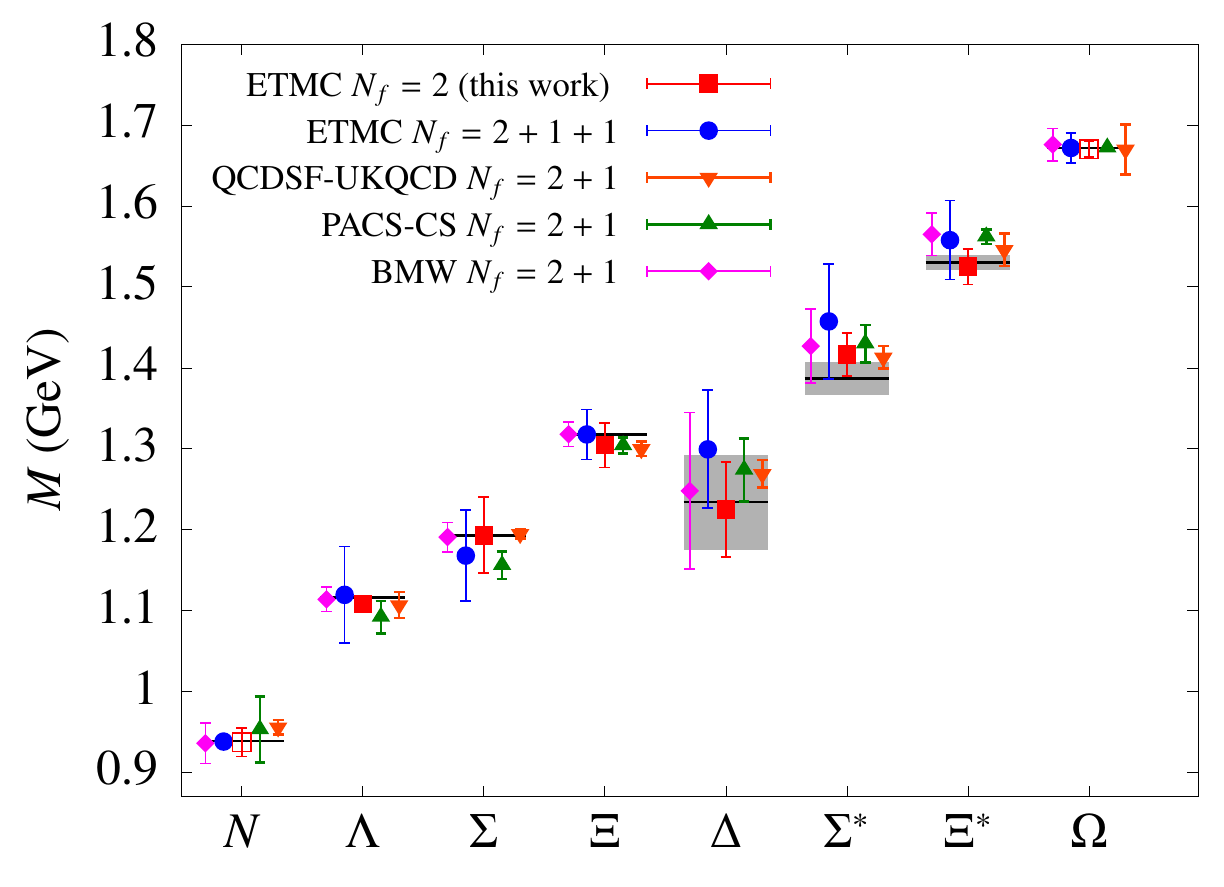}}
\caption{The octet and decuplet baryon masses obtained at the physical point and the experimental masses~\cite{Agashe:2014kda} shown by the horizontal bands. For most baryons the band is too small to be visible. The results of this work are shown with the red squares. The open squares in our results denote that the given mass was used as input. We additionally show the results extrapolated to the physical point from other lattice calculations for comparison~\cite{Alexandrou:2014sha,Aoki:2008sm,Bietenholz:2011qq,Durr:2008zz}. The results from ETMC~\cite{Alexandrou:2014sha} and BMW~\cite{Durr:2008zz} are also continuum extrapolated. More details are described in the text. The symbol notation is given in the legend of the figure.}
\label{Fig:spectrum_810}
\end{figure}

Regarding the octet and decuplet baryons, we compare with the results from the PACS-CS collaboration, obtained from $N_f=2+1$ non-perturbatively $\mathcal{O}(a)$ improved clover fermions on a lattice of spatial length of $2.9$ fm and a value of lattice spacing $a=0.09$ fm~\cite{Aoki:2008sm}. In addition, we compare with QCDSF-UKQCD results from Ref.~\cite{Bietenholz:2011qq}, using $N_f=2+1$ SLiNC configurations. The Budapest-Marseille-Wuppertal (BMW) collaboration have also obtained the strange baryon spectrum using tree level improved 6-step stout smeared $N_f=2+1$ clover fermions and a tree level Symanzik improved gauge action~\cite{Durr:2008zz} at $a=0.065,\;0.085$ and $0.125$ fm.
In \fig{Fig:spectrum_810} we show the masses for the octet and decuplet baryons using our physical ensemble, where we compare with the experimental values~\cite{Agashe:2014kda}, as well as with the results from other lattice QCD calculations. For our values we show the total error obtained by adding the statistical and systematic errors in quadrature. From the rest of the lattice calculations, only those from the ETMC~\cite{Alexandrou:2014sha} and BMW~\cite{Durr:2008zz} collaborations are continuum extrapolated. As can be seen, there is a good agreement among all lattice results. In particular, the results of this work computed directly at the physical point, although at finite value of the lattice spacing,  are in agreement with lattice QCD data that have been extrapolated to the continuum limit, indicating that cut-off effects are small. In addition they are in perfect agreement with experiment. We would like to point out that the large errors on our previous $N_f=2+1+1$ results are due to the systematic error arising from the chiral extrapolation.

A large number of groups have obtained partly or fully the charm baryon spectrum. The authors of Ref.~\cite{Na:2007pv,Na:2008hz} calculated the charm baryon spectrum using gauge configurations of the MILC collaboration with three degenerate flavours of Asqtad staggered sea quarks at three values of the lattice spacing, namely $a=0.09$, $0.12$ and $0.15$~fm. In Ref.~\cite{Briceno:2012wt} the charm baryon spectrum was obtained using the highly improved $N_f=2+1+1$ staggered  quark (HISQ) action at the sea, Wilson clover-improved light and strange fermions, and a relativistic heavy-quark action for the charm quark. Three lattice spacing values, $a=0.06$, $0.09$ and $0.12$~fm, were used and the continuum limit has been taken. In Ref.~\cite{Liu:2009jc} the masses of the singly charmed baryons were calculated, using domain wall fermions for the valence light and strange quarks, and the relativistic Fermilab action  for the valence charm quark, on  asqtad staggered sea quarks with a lattice spacing value of $a=0.12$~fm. More recent results include those from the PACS-CS collaboration, which obtained results directly at the physical point, using the relativistic heavy quark action on $N_f=2+1$ clover fermion configurations with the light and strange quarks tuned to their physical masses, a lattice spacing of $a=0.09$~fm and a spatial length of $L=2.9$ fm \cite{Namekawa:2013vu}. In Ref.~\cite{Brown:2014ena}  the charm baryon spectrum was computed using $N_f=2+1$ domain-wall fermions and a relativistic heavy-quark action for the charm quark. Two values of the lattice spacing, $a = 0.085$ and $0.112$~fm and seven values of  the pion mass were employed, and chiral and continuum extrapolations were performed. In addition, the Hadron Spectrum Collaboration (HSC) obtained results on the doubly charmed baryons from gauge ensembles using the tree-level Symanzik-improved gauge action and  clover fermions using an anisotropic lattice with the lattice spacing in the temporal direction  $a_t = 0.035$~fm and in the spatial directions $a_s=0.12$~fm, at a single pion mass of $m_\pi = 390$~MeV~\cite{Padmanath:2015jea}. Finally, the RQCD group~\cite{Bali:2015lka} has calculated the singly and doubly charmed baryon spectrum from $N_f=2+1$ non-perturbatively improved Wilson-clover fermions in a pion mass range of $m_\pi = 260\sim 460$~MeV and $a=0.075$~fm.

\begin{figure}[!ht]
\center
{\includegraphics[width=0.6\textwidth]{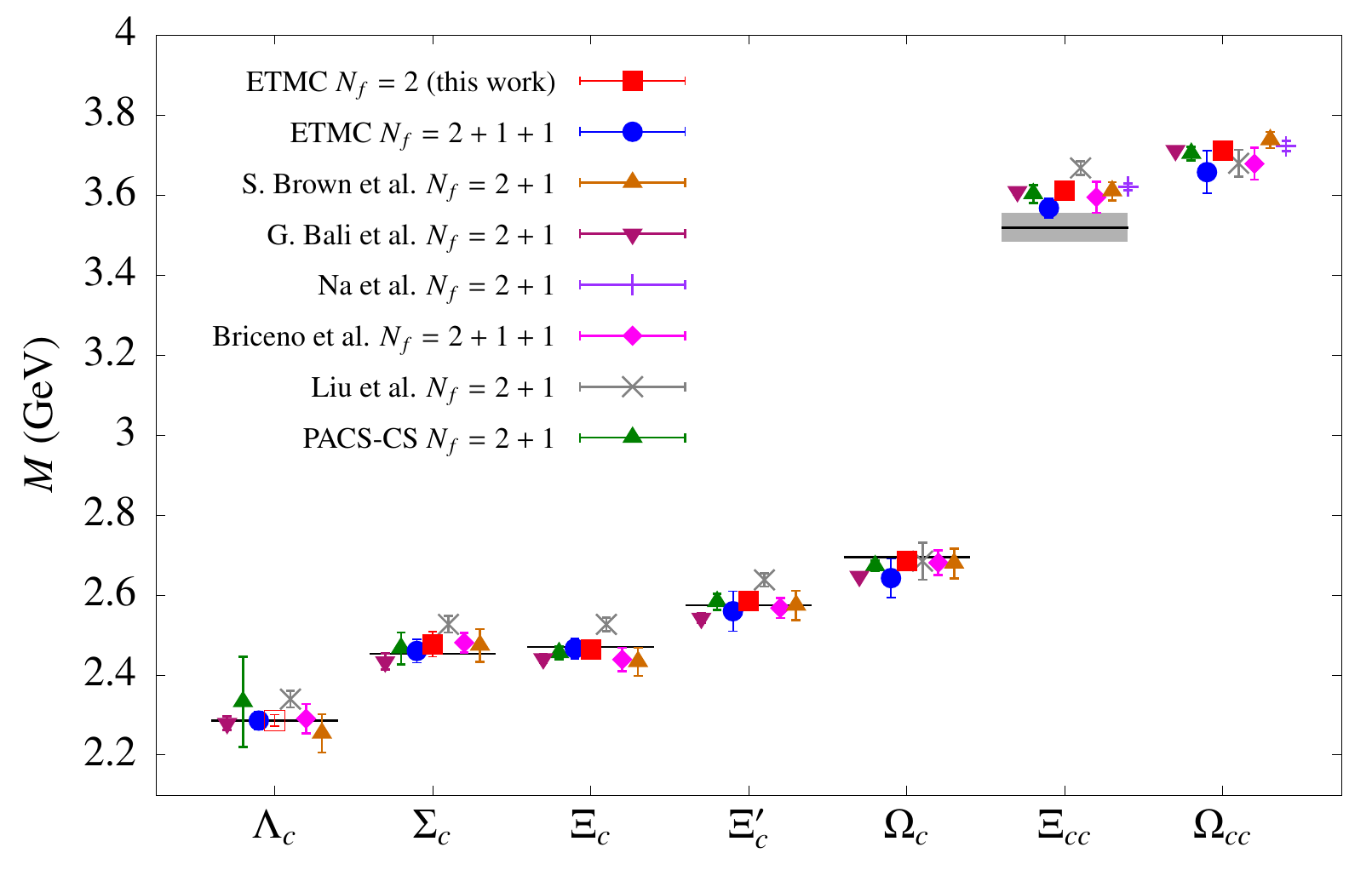}}
\caption{The masses of spin-1/2 charm baryons from this work (red squares) compared with the results extrapolated at the physical point from a number of other lattice calculations~\cite{Alexandrou:2014sha,Na:2007pv,Na:2008hz,Briceno:2012wt,Liu:2009jc,Brown:2014ena,Bali:2015lka}. The results from PACS-CS~\cite{Namekawa:2013vu} are obtained directly at the physical point. The results from ETMC~\cite{Alexandrou:2014sha}, R. A. Briceno et. al.~\cite{Briceno:2012wt} and Brown et. al.~\cite{Brown:2014ena} are also continuum extrapolated. The $\Lambda_c^+$ mass in our results was used as input, hence the open symbol. The experimental values~\cite{Agashe:2014kda}, wherever available, are shown with the horizontal bands. Details are given in the text. The symbol notation is given in the legend of the figure.}
\label{Fig:spectrum_charm12}
\end{figure}

\begin{figure}[!ht]
\center
{\includegraphics[width=0.6\textwidth]{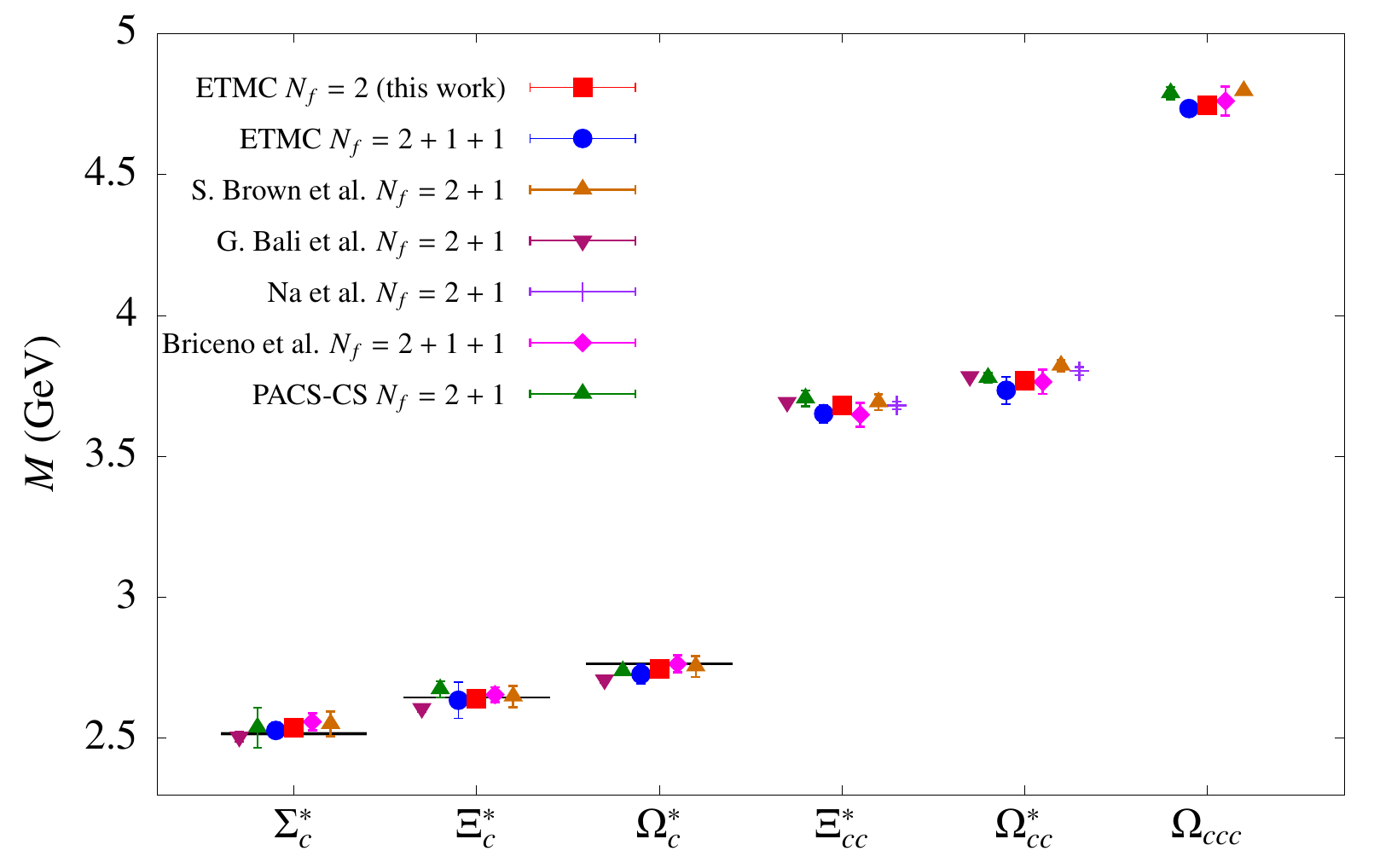}}
\caption{The masses of spin-3/2 charm baryons from this work (red squares) compared a number of other lattice calculations and with experiment, wherever available. The notation is as in~\fig{Fig:spectrum_charm12}. Details are given in the text.}
\label{Fig:spectrum_charm32}
\end{figure}

In Figs.~\ref{Fig:spectrum_charm12} and~\ref{Fig:spectrum_charm32} we illustrate the lattice QCD results mentioned above for the spin-1/2 and spin-3/2 charmed baryons, respectively, omitting the results from Ref.~\cite{Padmanath:2015jea} that were not extrapolated to the physical point. As in the octet and decuplet case, the error bar in our results denotes the statistical and systematic errors added in quadrature, however in most cases it is too small to be visible. The first important point to note is that there is an overall agreement among  the  lattice results, despite the fact that the continuum limit is not performed by all collaborations. This is a good indication that cut-off effect are small as compared to the statistical uncertainties, for the lattice spacings and improved actions used, which for the charm sector is a rather notable outcome. The second important point is that our results show perfect agreement with the experimental values even though the continuum limit has not been performed. This corroborates that cut-off effects are small for our action.  Only the mass of the doubly charmed $\Xi_{cc}$  is consistently overestimated by all the lattice results by $70 \sim 90$~MeV ($\sim 3\%$), which is yet to be confirmed by other experiments besides the SELEX measurement. Given this agreement,  lattice QCD can provide a rather robust prediction for  the $\Omega_{cc}$, $\Xi_{cc}^*$, $\Omega_{cc}^*$ and $\Omega_{ccc}$ masses that have not yet been measured experimentally.

Using our results we find the following values

\be
m_{\Omega_{cc}}   = 3.711(5)(30)\,{\rm GeV} \;,\; m_{\Xi_{cc}^*}   = 3.682(10)(26)\,{\rm GeV}\;,\;    m_{\Omega^*_{cc}} = 3.770(6)(30)\,{\rm GeV} \;,\; m_{\Omega_{ccc}}= 4.746(4)(32) \,{\rm GeV}\;,
\ee
where the first error is statistical and the second is the systematic due to the tuning of the strange and charm quark masses.

\section{Conclusions}\label{sec:conclusions}

Using an ensemble of $N_f=2$ twisted mass clover-improved fermions with physical values of the light quarks we compute the masses of the low-lying hyperon and charmed baryons. The strange and charm quarks are introduced as Osterwalder-Seiler fermions and their masses are tuned to reproduce the masses of the $\Omega^-$ and the $\Lambda_c^+$ baryons, respectively. The renormalized strange and charm quark masses are found to be 108.6(2.2)~MeV and 1392.6(23.5)~MeV, respectively, in the $\overline{\rm MS}$ scheme at 2~GeV at this value of the lattice spacing. Within one standard deviation, they are in agreement with other lattice QCD determinations. 

By having simulations with physical values of the quark masses we avoid chiral extrapolations, which in our previous studies were responsible for the largest systematic errors in our results. The large uncertainty in using chiral fits  is  reflected in the value we extract for the nucleon $\sigma_{\pi N}$ term using the Feynman-Hellmann theorem. The value we obtain from lowest and next-to-lowest chiral perturbation theory differ by almost 20\%. Both values are higher as compared to the recent values extracted using the direct approach where one computes the three point function of the scalar operator. Due to the large chiral extrapolation error, however, the two determination differ by one standard deviation. Nevertheless, given the fact that recent phenomenological analyses~\cite{Hoferichter:2012wf,Alarcon:2011zs,Pavan:2001wz} give rise to a larger value, more compatible with the one we find using the Feynman-Hellmann theorem, one needs to further examine the systematic errors involved in both determinations.
  
One of the disadvantages of the twisted mass formulation is that it breaks explicitly isospin symmetry at finite lattice spacing. In this work, we compute the isospin mass splitting in the baryon multiplets. In all cases the splitting is reduced by the inclusion of the clover term and in most cases the mass splitting is consistent with zero even for this rather coarse lattice spacing of 0.0938~fm. In particular, we find that for the spin-3/2 multiplets the mass splitting is consistent with zero. We find a mass splitting on the $\Sigma$ and $\Xi$ multiplets, which amounts to about 3\% of their masses. Small non-zero splitting is also found for the $\Xi_c$ multiplets. The splittings are taken as an additional systematic error in these cases.

Comparing our results with the experimental values wherever known we find perfect agreement, which allows us to predict the yet unmeasured masses of the doubly and triply charmed baryons. For the $\Xi_{cc}$ we find a mass of 3.606(11)(8)~GeV, which is higher by one standard deviation as compared 
with the value of 3.519~GeV  measured by the SELEX collaboration. Our prediction for the mass of the $\Xi_{cc}^*$ is 3.682(10)(26)~GeV, for the  $\Omega^+_{cc}$ is 3.711(5)(30)~GeV, for $\Omega^{*+}_{cc}$ 3.770(6)(30)~GeV and for  $\Omega^{++}_{ccc}$ 4.746(4)(32)~GeV.

\section*{Acknowledgments}

 We would like to thank all members of the ETMC for the many valuable and constructive discussions and the very fruitful collaboration that took place during the development of this work.  We acknowledge funding from the European Union's Horizon 2020 research and innovation programme under the Marie Sklodowska-Curie grant agreement No 642069.
The project used computer time granted by the John von Neumann Institute for Computing (NIC) on the JUROPA (now JURECA) system under the project \texttt{ecy00} at the J\"ulich Supercomputing Centre as well as by the Swiss Supercomputing Center CSCS under projects \texttt{s540} and \texttt{s625} and the Cyprus Institute on the Cy-Tera machine (project \texttt{lspro113s1}), under the Cy-Tera project NEA $\Upsilon\Pi$O$\Delta$OMH/$\Sigma$TPATH/0308/31. We thank the staff members of these computing centres for their technical advice and support. C.K. received partial support by the project GPUCW (T$\Pi$E/$\Pi\Lambda$HPO/0311(BIE)/09), which is co-financed by the European Regional Development Fund and the Republic of Cyprus through the Research Promotion Foundation.

%=============================================================
%=============================================================
%====== BIBLIOGRAPHY
\clearpage
\bibliographystyle{./apsrev}               % Style for bibliogrpahy
\bibliography{references}

%=============================================================
%=============================================================
%====== A P P E N D I C E S
\clearpage
\appendix
\begin{center}
{\bf Appendix A: Interpolating fields for baryons}\label{app:int_fields}
\end{center}

In the following tables we give the interpolating fields for baryons used in this work.
The sorting is in correspondence with \fig{fig:spin12_32}.
Throughout, $C$ denotes the charge conjugation matrix and the transposition sign refers to spinor indices which are suppressed.

\begin{table}[!ht]
\begin{center}
\renewcommand{\arraystretch}{1.2}
\renewcommand{\tabcolsep}{5.5pt}
\makebox[\textwidth]{%
\begin{tabular}{c|c|c c c c c}
\hline
\hline
\multirow{2}{*}{Charm} & \multirow{2}{*}{Strange} &  \multirow{2}{*}{Baryon} & Quark & \multirow{2}{*}{Interpolating field}  & \multirow{2}{*}{$I$} &   \multirow{2}{*}{$I_z$}   \\
                             &                         &                          & content  &                                     &                      &           \\
\hline
\renewcommand{\arraystretch}{1.6}
\renewcommand{\tabcolsep}{5.5pt}
%------------
\multirow{3}{*}{$c=2$}&   \multirow{2}{*}{$s=0$}                &$\Xi_{cc}^{++}$        & ucc &   $    \cone{c}{u}{c} $ & 1/2 & +1/2 \\
                                &                            &$\Xi_{cc}^{+}$        &  dcc & $    \cone{c}{d}{c} $ & 1/2 & -1/2 \\
\cline{2-7}
                         &$s=1$                 &$\Omega_{cc}^{+}$        & scc &   $    \cone{c}{s}{c} $  & 0 & 0\\
\hline\hline
%------------
\multirow{9}{*}{$c=1$}& \multirow{3}{*}{$s=0$}          &$\Sigma_c^{++}$  &  uuc & $    \cone{u}{c}{u} $ & 1 & +1 \\
                &               &$\Sigma_c^{+}$   & udc & $   \reci{\sqrt{2}} \eps \left[ \con{u}{c}{d} + \con{d}{c}{u} \right] $ & 1 & 0 \\
                        &               &$\Sigma_c^{0}$  &   ddc & $    \cone{d}{c}{d} $ & 1 & -1 \\
\cline{2-7}
        & \multirow{2}{*}{$s=1$}         &$\Xi_c^{\prime +}$  &  usc & $   \reci{\sqrt{2}} \eps \left[ \con{u}{c}{s} + \con{s}{c}{u} \right] $ & 1/2 & +1/2 \\
                                         &              &$\Xi_c^{\prime 0}$  &   dsc &$   \reci{\sqrt{2}} \eps \left[ \con{d}{c}{s} + \con{s}{c}{d} \right] $ & 1/2 & -1/2 \\
\cline{2-7}
                  & $s=2$       &$\Omega_c^{0}$        & ssc &   $    \cone{s}{c}{s} $ & 0 & 0\\
\cline{2-7}
 & $s=0$ & $\Lambda_c^+$  &     udc      & $   \reci{\sqrt{6}} \eps \left[ 2\con{u}{d}{c} + \con{u}{c}{d} - \con{d}{c}{u} \right] $   & 0 & 0   \\
\cline{2-7}
%& \multirow{2}{*}{$s=1$}               &$\Xi_c^+$      & usc & $   \cone{u}{s}{c} $ & 1/2 & +1/2 \\
% &       &$\Xi_c^0$   & dsc & $   \cone{d}{s}{c} $ & 1/2 & -1/2 \\
& \multirow{2}{*}{$s=1$}                &$\Xi_c^+$      & usc & $   \reci{\sqrt{6}} \eps \left[ 2\con{s}{u}{c} + \con{s}{c}{u} - \con{u}{c}{s} \right] $ & 1/2 & +1/2 \\
 &        &$\Xi_c^0$   & dsc & $   \reci{\sqrt{6}} \eps \left[ 2\con{s}{d}{c} + \con{s}{c}{d} - \con{d}{c}{s} \right] $ & 1/2 & -1/2 \\
\hline\hline
%------------
\multirow{8}{*}{$c=0$}& \multirow{2}{*}{$s=0$}     & \emph{p}     &     uud           &    $\cone{u}{d}{u} $      & 1/2 & +1/2     \\
        &               & \emph{n}     & udd       & $   \cone{d}{u}{d} $      & 1/2 & -1/2      \\
\cline{2-7}
& \multirow{4}{*}{$s=1$}                &$\Lambda$     & uds     &      $   \reci{\sqrt{6}} \eps \left[ 2\con{u}{d}{s} + \con{u}{s}{d} - \con{d}{s}{u} \right] $ & 0 & 0 \\
\cline{3-7}
                &       &$\Sigma^{+}$  &  uus    & $    \cone{u}{s}{u} $     & 1 & +1\\
      &         &$\Sigma^{0}$  & uds     & $   \reci{\sqrt{2}} \eps \left[ \con{u}{s}{d} + \con{d}{s}{u} \right] $& 1 & 0 \\
           &   &$\Sigma^{-}$  &   dds   & $    \cone{d}{s}{d} $     & 1 & -1\\
\cline{2-7}
& \multirow{2}{*}{$s=2$}    &$\Xi^{0}$      & uss    &   $    \cone{s}{u}{s} $    & 1/2 & +1/2 \\
                &   &$\Xi^{-}$     &  dss   & $    \cone{s}{d}{s} $       & 1/2 & -1/2\\
\hline\hline
\end{tabular}
}
\end{center}
\caption{Interpolating fields and quantum numbers for the $20^\prime$-plet of spin-1/2 baryons.}
\label{spin12_tab}
\end{table}
%
%\vspace*{1.5cm}
\clearpage
\begin{table}[!ht]
\begin{center}
\renewcommand{\arraystretch}{1.2}
\renewcommand{\tabcolsep}{5.5pt}
\makebox[\textwidth]{%
\begin{tabular}{c|c|c c c c c}
\hline
\hline
\multirow{2}{*}{Charm} & \multirow{2}{*}{Strange} &  \multirow{2}{*}{Baryon} & Quark & \multirow{2}{*}{Interpolating field}  & \multirow{2}{*}{$I$} &   \multirow{2}{*}{$I_z$}   \\
                             &                         &                          & content  &                                     &                      &           \\
\hline
\renewcommand{\arraystretch}{1.6}
\renewcommand{\tabcolsep}{5.5pt}
%------------
        $c=3$ &  $s=0$  &$\Omega_{ccc}^{++}$        & ccc &   $     \conme{c}{c}{c} $ & 0 & 0 \\
\hline\hline
%------------
 \multirow{3}{*}{$c=2$}    & \multirow{2}{*}{$s=0$}     &$\Xi_{cc}^{\star ++}$        & ucc &  $   \reci{\sqrt{3}} \eps \left[ 2\conm{c}{u}{c} + \conm{c}{c}{u} \right] $ & 1/2 & +1/2\\
                                &               &$\Xi_{cc}^{\star +}$        &  dcc & $   \reci{\sqrt{3}} \eps \left[ 2\conm{c}{d}{c} + \conm{c}{c}{d} \right] $ & 1/2 & -1/2\\
\cline{2-7}
                                &$s=1$  &$\Omega_{cc}^{\star +}$        & scc &   $   \reci{\sqrt{3}} \eps \left[ 2\conm{c}{s}{c} + \conm{c}{c}{s} \right] $ & 0 & 0\\
\hline\hline
%------------
\multirow{6}{*}{$c=1$}&   \multirow{3}{*}{$s=0$}        &$\Sigma_c^{\star ++}$  & uuc &   $   \reci{\sqrt{3}} \eps \left[ \conm{u}{u}{c} + 2\conm{c}{u}{u} \right] $ & 1 & +1 \\
                         &                          &$\Sigma_c^{\star +}$                  & udc   & $   \sqrt{\frac{2}{3}} \eps \left[ \conm{u}{d}{c} + \conm{d}{c}{u}+ \conm{c}{u}{d} \right] $ & 1 & 0 \\
                        &                       &$\Sigma_c^{\star 0}$  &   ddc        &   $     \reci{\sqrt{3}} \eps \left[ \conm{d}{d}{c} + 2\conm{c}{d}{d} \right] $ & 1 & -1 \\

\cline{2-7}
       &          \multirow{2}{*}{$s=1$}        &$\Xi_c^{\star +}$      & usc & $   \sqrt{\frac{2}{3}} \eps \left[ \conm{u}{s}{c} + \conm{s}{c}{u}+ \conm{c}{u}{s} \right] $  & 1/2 & +1/2\\
                 &                      &$\Xi_c^{\star 0}$   & dsc & $   \sqrt{\frac{2}{3}} \eps \left[ \conm{d}{s}{c} + \conm{s}{c}{d}+ \conm{c}{d}{s} \right] $  & 1/2 & -1/2\\
\cline{2-7}
                  &$s=2$        &$\Omega_c^{\star 0}$        & ssc & $   \reci{\sqrt{3}} \eps \left[ 2\conm{s}{c}{s} + \conm{s}{s}{c} \right] $ & 0 & 0  \\
\hline\hline
%------------
\multirow{10}{*}{$c=0$}&   \multirow{4}{*}{$s=0$}               & $\Delta^{++}$  & uuu & $    \conme{u}{u}{u} $   & 3/2 & +3/2         \\
        &       & $\Delta^{+}  $  &  uud & $   \reci{\sqrt{3}} \eps \left[ 2\conm{u}{d}{u} + \conm{u}{u}{d} \right] $& 3/2 & +1/2         \\
        &       & $\Delta^{0}  $  &  udd & $   \reci{\sqrt{3}} \eps \left[ 2\conm{d}{u}{d} + \conm{d}{d}{u} \right] $ & 3/2 & -1/2        \\
        &       & $\Delta^{-}$  &       ddd & $    \conme{d}{d}{d} $   & 3/2 & -3/2         \\
\cline{2-7}
   & \multirow{3}{*}{$s=1$}              &$\Sigma^{\star +}$  & uus &   $   \reci{\sqrt{3}} \eps \left[ \conm{u}{u}{s} + 2\conm{s}{u}{u} \right] $ & 1 & +1 \\
    &   &$\Sigma^{\star 0}$                & uds   & $   \sqrt{\frac{2}{3}} \eps \left[ \conm{u}{d}{s} + \conm{d}{s}{u}+ \conm{s}{u}{d} \right] $& 1 & 0 \\
         & &$\Sigma^{\star -}$  &   dds        &   $     \reci{\sqrt{3}} \eps \left[ \conm{d}{d}{s} + 2\conm{s}{d}{d} \right] $ & 1 & -1\\
\cline{2-7}
%& \multirow{2}{*}{$s=2$}               &$\Xi^{\star 0}$        & uss &   $     \conme{s}{u}{s} $ & 1/2 & +1/2\\
%               &       &$\Xi^{\star -}$        &  dss & $     \conme{s}{d}{s} $ & 1/2 & -1/2\\
& \multirow{2}{*}{$s=2$}                &$\Xi^{\star 0}$        & uss &  $   \reci{\sqrt{3}} \eps \left[ 2\conm{s}{u}{s} + \conm{s}{s}{u} \right] $ & 1/2 & +1/2\\
                &       &$\Xi^{\star -}$        &  dss & $   \reci{\sqrt{3}} \eps \left[ 2\conm{s}{d}{s} + \conm{s}{s}{d} \right] $ & 1/2 & -1/2\\
\cline{2-7}
& $s=3$         & $\Omega^-$  & sss &$   \conme{s}{s}{s}$ &0&0 \\
\hline\hline
\end{tabular}
}
\end{center}
\caption{Interpolating fields and quantum numbers for the 20-plet of spin-3/2 baryons.}
\label{spin32_tab}
\end{table}

%==============================================================

\newpage

\begin{center}
{\bf Appendix B: Lattice results}\label{app:tables}
\end{center}

\begin{table}[h]
\begin{center}
\renewcommand{\arraystretch}{1.4}
\renewcommand{\tabcolsep}{5.8pt}
\begin{tabular}{c|c|c|c|c|c|c}
\hline\hline
 Volume & Statistics & $a\mu_l$ & $am_\pi$ & $m_\pi$ (GeV) & $am_N$ & $m_N$ (GeV) \\
\hline\hline
\multicolumn{7}{c}{$N_f=2+1+1$, $\beta=1.90$} \\
\hline
\multirow{3}{*}{$32^3\times64$} & 2960    &  0.0030   & 0.1240 &  0.2607  &  0.5239(87)  & 1.1020(183)  \\
 & 6224   &  0.0040   & 0.1414 &  0.2975  &  0.5192(112) & 1.0921(235)  \\
 & 1548    &  0.0050   & 0.1580 &  0.3323  &  0.5422(62)  & 1.1407(130)  \\
\hline
\multirow{3}{*}{$24^3\times48$} & 8368   &  0.0400   & 0.1449 &  0.3049  &  0.5414(84)  & 1.1389(176)  \\
 & 7664   &  0.0060   & 0.1728 &  0.3634  &  0.5722(48)  & 1.2036(101)  \\
 & 7184   &  0.0080   & 0.1988 &  0.4181  &  0.5898(50)  & 1.2407(104)  \\
 & 8016   &  0.0100   & 0.2229 &  0.4690  &  0.6206(43)  & 1.3056(90)   \\
\hline
$20^3\times 48$ & 2468    &  0.0040   & 0.1493 &  0.3140  &  0.5499(195) & 1.1568(410)  \\
\hline\hline
\multicolumn{7}{c}{$N_f=2+1+1$, $\beta=1.95$}\\
\hline
\multirow{4}{*}{$32^3\times64$} & 2892   &  0.0025   & 0.1068 &  0.2558  &  0.4470(59)  & 1.0706(141)  \\
 & 4204   &  0.0035   & 0.1260 &  0.3018  &  0.4784(48)  & 1.1458(114)  \\
 & 18576  &  0.0055   & 0.1552 &  0.3716  &  0.5031(16)  & 1.2049(39)   \\
 & 2084   &  0.0075   & 0.1802 &  0.4316  &  0.5330(42)  & 1.2764(100)  \\
\hline
$24^3\times 48$  & 937    &  0.0085   & 0.1940 &  0.4645  &  0.5416(50)  & 1.2970(121)  \\
\hline\hline
\multicolumn{7}{c}{$N_f=2+1+1$, $\beta=2.10$}\\
\hline
\multirow{3}{*}{$48^3\times96$} & 2424   &  0.0015   & 0.0698 &  0.2128  &  0.3380(41)  & 1.0310(125)  \\
 & 744    &  0.0020   & 0.0805 &  0.2455  &  0.3514(70)  & 1.0721(215)  \\
 & 904    &  0.0030   & 0.0978 &  0.2984  &  0.3618(68)  & 1.1038(208)  \\
\hline
 $32^3\times 64$ & 7620   &  0.0045   & 0.1209 &  0.3687  &  0.3944(26)  & 1.2032(79)   \\
\hline\hline
\multicolumn{7}{c}{$N_f=2$, $\beta=2.10$, $c_{\rm sw}=1.57551$}\\
\hline
 $48^3\times 96$ & 861200   &  0.0009   & 0.0621 &  0.1305  &  0.4436(11)  & 0.9321(36)   \\
\hline\hline
\end{tabular}
\caption{Values of the pion and nucleon masses for the $N_f=2+1+1$ ensembles with the associated statistical error that were used in the fits for the determination of the lattice spacing. Also included is the physical ensemble used in this work. The lattice spacings of Eqs.~\ref{eq:lat_spacing} and~\ref{eq:nf211_alat} were used for converting in physical units.}
\label{Table:nucleon_masses}
\end{center}
\vspace*{-.0cm}
\end{table} 

%==============================================================

\end{document}